\documentclass[iop,numberedappendix,appendixfloats]{emulateapj}

\usepackage{epstopdf}
\usepackage{graphics}
\usepackage{natbib}

\newcommand{\msun}{\mbox{$\rm M_{\odot}$}}

\newcommand{\lsun}{\mbox{L$_{\odot}$}}
\newcommand{\kms}{\mbox{km s$^{-1}$}}

\newcommand{\etal}[1]{{ et al.}~}
\def\kms{\ifmmode \hbox{km~s}^{-1}\else km~s$^{-1}$\fi}
\def\etal {{\it et al.}}
\def\deg      {{\ifmmode^\circ\else$^\circ$\fi} } 

\def\h2     {H$_2$}

\def\arcsec{\hbox{$^{\prime\prime}$}}

\shorttitle{Arp 220 and NGC 6240}
\shortauthors{Scoville \etal}

\begin{document}

\title{ALMA Imaging of HCN, CS and dust in Arp 220 and NGC 6240}
 \author{Nick Scoville\altaffilmark{1}, Kartik Sheth\altaffilmark{3}, Fabian Walter\altaffilmark{2}, Swarnima Manohar\altaffilmark{1}, Laura Zschaechner\altaffilmark{2}, Min Yun\altaffilmark{8}, Jin Koda\altaffilmark{15}, David Sanders\altaffilmark{5}, Lena Murchikova\altaffilmark{1}, Todd Thompson\altaffilmark{13,14}, Brant Robertson\altaffilmark{11}, Reinhard Genzel\altaffilmark{9}, Lars Hernquist\altaffilmark{6},  Linda Tacconi\altaffilmark{9}, Robert Brown\altaffilmark{10}, Desika Narayanan\altaffilmark{11},  Christopher C. Hayward\altaffilmark{7}, Joshua Barnes\altaffilmark{5},  Jeyhan Kartaltepe\altaffilmark{4},  Richard Davies\altaffilmark{9}, Paul van der Werf\altaffilmark{12},  Edward Fomalont\altaffilmark{10,3}}
 
\altaffiltext{}{}
\altaffiltext{1}{California Institute of Technology, MC 249-17, 1200 East California Boulevard, Pasadena, CA 91125}
\altaffiltext{2}{Max-Planck-Institut fur Astronomie, Konigstuhl 17, D-69117 Heidelberg, Germany}
\altaffiltext{3}{North American ALMA Science Center, National Radio Astronomy Observatory, 520 Edgemont Road, Charlottesville, VA 22901, USA}
\altaffiltext{4}{National Optical Astronomy Observatory, 950 North Cherry Avenue, Tucson, AZ 85719, USA}
\altaffiltext{5}{Institute for Astronomy, 2680 Woodlawn Dr., University of Hawaii, Honolulu, Hawaii, 96822}
\altaffiltext{6}{Harvard-Smithsonian Center for Astrophysics, 60 Garden Street, Cambridge, MA 02138, USA}
\altaffiltext{7}{TAPIR 350-17, California Institute of Technology, 1200 E. California Boulevard, Pasadena, CA 91125}
\altaffiltext{8}{Department of Astronomy, University of Massachusetts, Amherst, MA 01003}
\altaffiltext{9}{Max-Planck-Institut fur extraterrestrische Physik (MPE), Giessenbachstr., D-85748 Garching, Germany}
\altaffiltext{10}{National Radio Astronomy Observatory, 520 Edgemont Road, Charlottesville, VA 22901, USA}
\altaffiltext{11}{Department of Astronomy and Steward Observatory, University of Arizona, Tucson AZ 85721}
\altaffiltext{12}{Leiden Observatory, Leiden University, P.O. Box 9513, NL-2300 RA Leiden, The Netherlands}
\altaffiltext{13}{Department of Astronomy, The Ohio State University, 140 West 18th Avenue, Columbus, OH 43210, USA}
\altaffiltext{14}{Center for Cosmology and AstroParticle Physics, The Ohio State University, 191 West Woodruff Avenue, Columbus, OH 43210, USA}
\altaffiltext{15}{Department of Physics \& Astronomy, Stony Brook University, Stony Brook, NY 11794, USA}
\altaffiltext{}{}

\vfill
\eject

\begin{abstract}
We report ALMA Band 7 (350 GHz) imaging at 0.4 - 0.6\arcsec ~resolution and Band 9 (696 GHz) 
at $\sim$0.25\arcsec ~resolution of the luminous IR galaxies Arp 220 and NGC 6240.  The long wavelength dust continuum is used to estimate 
ISM masses for Arp 220 East, West and NGC 6240 of 1.9, 4.2 and 1.6$\times10^9$ \msun within 
radii of 69, 65 and 190 pc. The HCN emission was modeled to derive the emissivity distribution as a function of radius and the kinematics of 
each nuclear disk, yielding dynamical masses consistent with the masses and sizes derived from the dust emission. In Arp 220, the major dust and gas 
concentrations are at radii less than 50 pc in both counter-rotating nuclear 
disks. The thickness of the disks in Arp 220 
estimated from the velocity dispersion and rotation 
velocities are 10-20 pc and the mean gas densities 
are $n_{H_2} \sim10^5$ cm$^{-3}$ at R $< 50$ pc. We develop 
an analytic treatment for 
the molecular excitation (including photon trapping), yielding volume densities 
for both the HCN and CS emission with $n_{H_2} \sim2\times10^5$ cm$^{-3}$. 
The agreement of the mean density from the total mass and size with that required for excitation 
suggests that the volume is essentially filled with dense gas, i.e. it is not cloudy or 
like swiss cheese. 
\end{abstract}

\medskip 

 \keywords{galaxy evolution ISM: clouds --- galaxies: individual (Arp 220, NGC 6240) --- galaxies: active, starburst, interactions -- ISM: molecules}

\section{Introduction}

Numerous multi-wavelength observational and theoretical investigations have revealed the basic properties and evolutionary scenario 
of ultra-luminous infrared galaxies (ULIRGs). The first complete sample for the local universe was generated from the IRAS all sky  
survey, consisting of 12 galaxies with L$_{1-1000\mu m} > 10^{12}$ \lsun ~at $z < 0.1$, and follow-up ground-based optical imaging 
revealed virtually all of the sample to be merging galaxies or post merging systems \citep{san88,san96}. All are also gas-rich with 
molecular gas masses determined from CO to be $2 - 20\times10^9$ \msun \citep{san91}. For the ULIRGs, approximately equal numbers of galaxies have optical emission line ratios characteristic of star formation (SF) and AGN \citep{san88}. Although the ULIRG activity is rare at low redshift, these merging systems 
are likely much more prevalent during the epoch 
of peak cosmic star formation and nuclear activity at z $\sim1 - 4$ \citep{lefl05}.

Arp 220 (L$_{IR} = 2.5\times10^{12}$ \lsun, D$_L$ = 77 Mpc, D$_A$ = 74 Mpc, 361 pc arcsec$^{-1}$) and NGC 6240 (L$_{IR} = 0.9\times10^{12}$ \lsun, D$_L$ = 103 Mpc, D$_A$ = 98 Mpc, 475 pc arcsec$^{-1}$) are two of the most studied luminous IR galaxies. 
Both systems show double nuclei -- separated by $1.0$\arcsec (361 pc, Arp 220) and 1.5\arcsec (713 pc, NGC 6240). 
In Arp 220 the star formation rate must be $\sim240$ \msun~yr$^{-1}$ if the infrared luminosity is powered by star formation. In NGC 6240  
 both galactic nuclei show X-ray emission suggesting that AGN accretion contributes some of the luminosity \citep{kom03}. In NGC 6240 the molecular gas peak lies between the
two NIR and X-ray nuclei so it is likely that most of the IR luminosity originates from star formation (SFR $\sim70$ \msun ~yr$^{-1}$) rather than AGN accretion on the nuclei.
 
During the merging of gas-rich galaxies, the original ISM (presumably distributed in extended galactic-scale disks) sinks rapidly to the center of the merging system due to dissipation of kinetic energy in the shocked gas and torques generated by the offset stellar and gaseous bars \citep[e.g.][]{bar92}. The star formation rates in the ULIRGs are typically 10-100 times higher per unit mass of ISM compared to quiescent disk galaxies. The ULIRG-starburst activity is likely driven by the 
concentration of gas in the nuclear regions and dynamical compression of this gas by supersonic shocks. 
 Arp 220 has two nuclei separated by $\sim$300 pc and over 90\% of its bolometric luminosity emerges in the infrared. The two nuclei have counter-rotating disks of molecular gas as traced by CO emission \citep{sak99}. In NGC 6240, the molecular gas extends between the nuclei rather than being concentrated on the individual nuclei. 

The observations presented here were made with the ALMA array at 350 and 700 GHz (Bands 7 and 9). These data provide excellent resolution and sensitivity for imaging high excitation 
molecular gas tracers (HCN and CS) together with the optically thin long wavelength dust continuum. The dust continuum provides an independent and  linear probe of the 
overall ISM mass \citep{sco14}. 


\subsection{Arp 220 -- summary of prior observations}

Near infrared imaging of Arp 220 reveals two nuclei separated by 1\arcsec;  the West nucleus shows 
greatly increased extinction to the south 
, modeled as a dense dust disk tilted to the line of sight
\citep{sco98}. CO imaging at 0.5\arcsec~ resolution detects the counter-rotating disks in both nuclei with radii $\sim100$ pc, dynamical 
masses of $2\times10^9$ \msun ~for each disk and $\sim50$\% gas mass fractions \citep{sco97,sak99,dow07}. The inferred visible extinctions
perpendicular to the disks are A$_V \sim 500 - 2000$mag, precluding the use of optical/near infrared tracers of the star formation 
in the nuclei. Near infrared integral field spectroscopy of the stars at larger radii indicates a significant population with age only 10 Myr \citep{eng11}.  Much of the nuclear disk 
ISM is very dense ($>10^{4-5}$cm$^{-3}$) and at high temperatures  ($T > 75K$) compared to normal Galactic molecular gas. This is based on the strong HCN \& HCO$^+$ line emission observed with single dish telescopes and the high 
CO brightness and dust temperatures $\sim75 - 170$ K \citep{mat09,gre09,sak99,dow07,ran11,wil14}.  PCygni absorption line profiles are seen in HCO$^+$ (J = 3 - 2 and 4 - 3) with SMA, which \cite{sak09} interpret
 as an outflow from the inner radii of the nuclear disks. HNC (J = 3 - 2) emission 
has also been imaged at $0.3$\arcsec ~resolution with SMA, but the emission is bright only in the western nucleus \citep{aal09}.

\subsection{NGC 6240 -- summary of prior observations}
In the infrared, NGC 6240 shows two nuclei separated by $\sim2$ arcsec (700 pc) 
with high reddening in the region between them \citep[based on the 2.2/1.6 $\mu$m color gradients -- ][]{sco00}. Each of the nuclei exhibits X-ray emission 
suggestive of supermassive black hole accretion \citep{kom03}. High 
resolution CO (J = 2-1) imaging shows the molecular gas peaking in this 'overlap region' \citep{bry99,tac99,eng10}. The 
gas concentration between the nuclei can be modeled as a rotating disk with dynamical mass $7\times10^9$ \msun~ and a molecular 
gas mass fraction of $\sim50$\% \citep{tac99}. \cite{ion07} finds the CO (3 - 2) emission extended on a scale of 4 kpc but the HCO$^+$ (4 - 3) emission is concentrated between the two nuclei.  
Near infrared spectroscopy of NGC 6240 indicates a significant starburst (SB) population less than 20 Myr old \citep{tec00,eng10}.

\subsection{Probing the Nuclear Structure}

Although the global understanding of merger evolution is well developed, major aspects of the structure and evolution of the massive gas-rich nuclear disks remain poorly constrained by observations and only 
crudely understood in terms of the physical processes. Specifically, the radial mass and star formation distributions, the physical conditions (density, temperature and cloud structures) and the dynamical 
evolution of the disks (due to accretion and feedback from SB and AGN activities) are undetermined. And yet, all of these are vital to understanding the final evolution of the mergers and the resulting SB properties (SF efficiency, distribution and feedback), AGN fueling and the feedback of processed material to the circum-galactic environment.  

Here, we present ALMA imaging at 350 GHz (850 $\mu$m) of Arp 220 and NGC 6240 at 0.3 - 0.6\arcsec~resolution. The observations include both the 
dust continuum, which is optically thin at long wavelengths and the emission lines of high excitation molecular gas tracers (HCN, CS and CN). 
The HCN and CS emissions trace higher excitation molecular gas than is probed by the CO lines, since they require high column densities (and/or volume densities) 
to maintain population in the excited levels from which the emission originates. Thus, they provide better access to the gas closest to the centers of SB and AGN activity (as opposed
to the extended molecular gas producing the CO emission). 

\bigskip
We describe the observations and data reduction in \S \ref{obs} and summarize the 
results in \S \ref{results}. In \S \ref{dust_mass}, we use the dust emission to estimate the masses of ISM in each source.
In \S \ref{nuclear_disks} and Appendix \ref{app}, 
we model the nuclear disk dynamics and HCN emissivity distributions in Arp 220 and NGC 6240, doing a maximum likelihood fit to obtain a best match to the observed HCN emission line profiles. 
This dynamical modeling yields total masses consistent with those derived from the dust continuum emission. 

\section{ALMA Observations and Data Reduction}\label{obs}

The ALMA Cycle 0 observations for project \#2011.0.00175.S 
were obtained in 2012. For both Band 7 and
Band 9 observations the correlator was configured in the time division
mode (TDM) with 4 spectral windows. Each window had a full bandwidth of 1875 MHz covered by
31.25 MHz resolution spectral channels.  Three of the four windows in Band 7 were
configured to observe spectral lines at rest frequencies of 342.882
(CS 7 - 6) 
and 354.505 GHz (HCN J=4 - 3), while the
fourth spectral window was centered at 340.974 GHz to image the
dust continuum emission.  For the  Band 9 observations, we configured the correlator to
map two spectral lines at rest frequencies of 708.877 (HCN 8-7) and
713.341 GHz (HCO$^+$ 8-7) and used the two remaining spectral windows to
image the continuum emission -- these latter windows were centered at rest frequencies 707.3 and 711.0 GHz.  
The adopted redshifts were $z= 0.01812$ and $0.02448$ for Arp 220 and NGC 6240, respectively. For Band 9, we only present the continuum data here. 
[The Band 9 line emission data are not discussed further here and will be presented in a separate publication.]

All observations were done in the most extended
configuration available in Cycle 0. For this telescope configuration, good flux recovery 
is expected out to scales of
$\sim3$\arcsec~and more extended emission will be partially resolved out. The spatial scales of the high excitation gas is confined to $<$3\arcsec ~and
observations with the ALMA compact telescope configuration were not necessary. The CO and dust continuum emission from the 
larger galaxies is likely to be significantly resolved out at 3\arcsec scales \citep{sco97}. 
Most of the data were taken with configurations of 16 -- 21 antennas, except one execution block which had 25 antennas.  The total integration time
for the Band 7 observations for NGC 6240 and Arp 220 was 140 minutes for each source (excluding calibrations), whereas for the Band 9
observations, the total on-source time was 32 minutes for Arp 220
and 75 minutes for NGC 6240.  

Following the delivery of data products, the data were re-reduced and imaged
using the Common Astronomy Software Applications package (CASA).
Modest self-calibration was done to improve the dynamic range.  We measured
the noise from the emission free regions of the map before subtraction of the continuum.  The 1$\sigma$ (rms) sensitivities are as follows: for Arp 220
in Band 7, we measure an rms noise of $1.1\rm ~mJy beam^{-1}$ in 13~\kms ~channels in the
upper sideband spectral windows (i.e. at a rest frequency of $\sim$
354 GHz) and $ 1.7\rm ~mJy beam^{-1}$  in 13 \kms channels in the lower sideband spectral
windows (i.e. at a rest frequency of 342 GHz). For NGC 6240 we
have a noise level of $\sigma \simeq 0.5 \rm ~mJy~ beam^{-1}$ and $0.6
 \rm ~mJy~ beam^{-1}$ in the upper and lower sidebands in 13 \kms ~ channels.  The
poorer sensitivity in the Arp 220 observations is largely due to the lower elevation of 
Arp 220 and shorter integration times.

\begin{figure*}[t]
\bigskip
\plotone{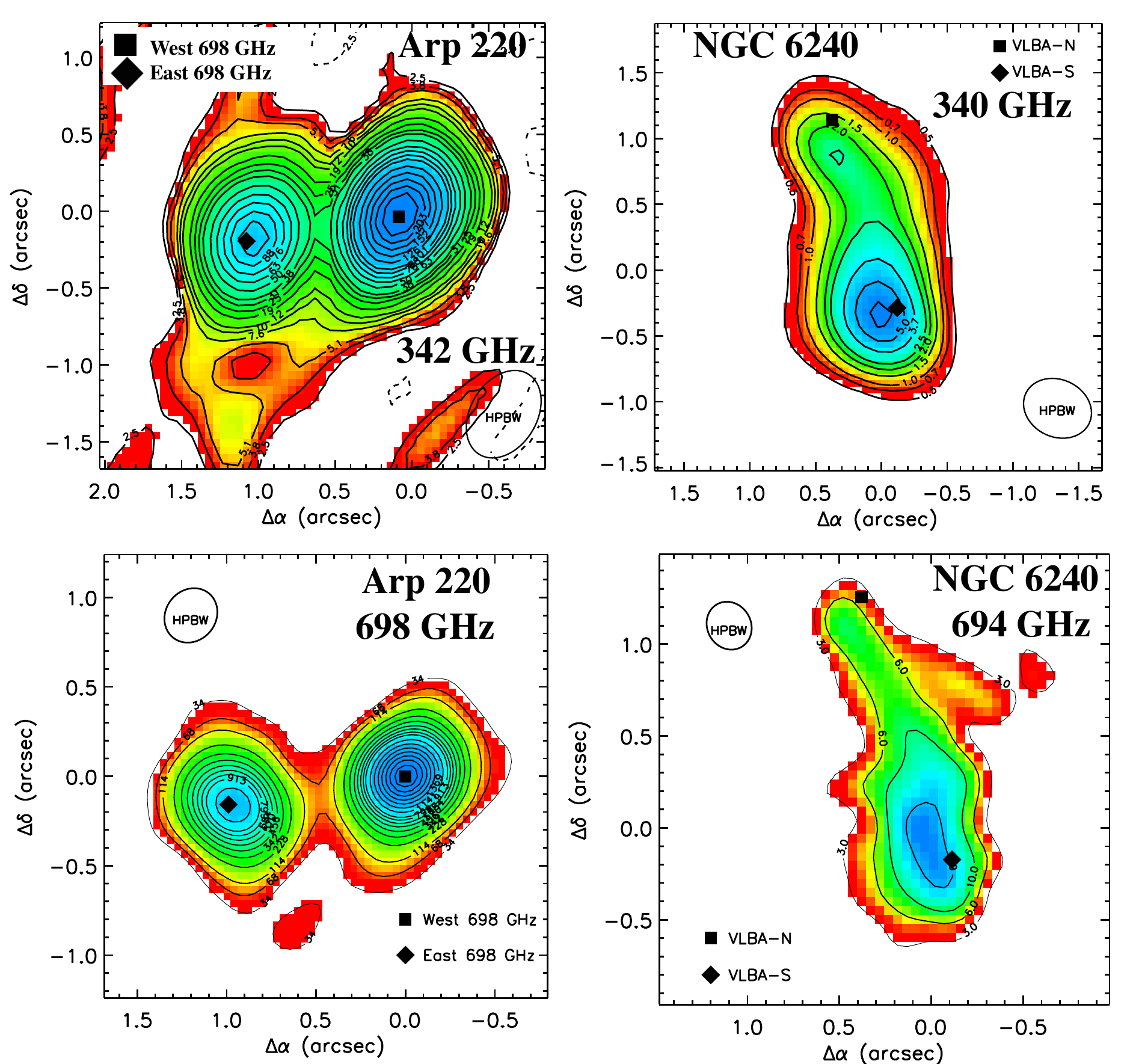}
\caption{Continuum images of Arp 220 and NGC 6240. In the Arp 220 images, the square and diamond markers indicate the West and East continuum Gaussian-fit peaks at 698 GHz (see Table \ref{fits}) and the coordinate offsets 
are relative the West continuum peak. In the NGC 6240 images, the square and diamond markers indicate the North and South VLBA continuum peaks \citep[see][]{max07}. Labels on the contours are in mJy beam$^{-1}$. Contours are at -2 (dashed), 2, 3, 4, 6, 8, 10, 15, 20, 25, 30, 40,... 80, 100, ...$ \times \sigma$
where $\sigma = 1.26 \rm ~and~0.248$ mJy beam$^{-1}$,respectively for Arp 220 at 342 GHz and NGC 6240 at 340 GHz. For Arp 220 at 698 GHz and NGC 6240 at 694 GHz, contours are at  -3 (dashed), 3, 6, 10,  20,  ... 100, 120, 140, ... $ \times \sigma$
where $\sigma = 11 \rm ~and~1$ mJy beam$^{-1}$. The beam sizes are shown (Arp 220 Band 7: $0.60-0.52\times0.42-0.39$\arcsec ~and Band 9: $0.32\times0.28$\arcsec; NGC 6240 Band 7: $0.54\times045$\arcsec ~and Band 9: $0.27\times0.24$\arcsec).} 
\label{continuum} 
\end{figure*}

\begin{figure}[t]
 \epsscale{1.}
\plotone{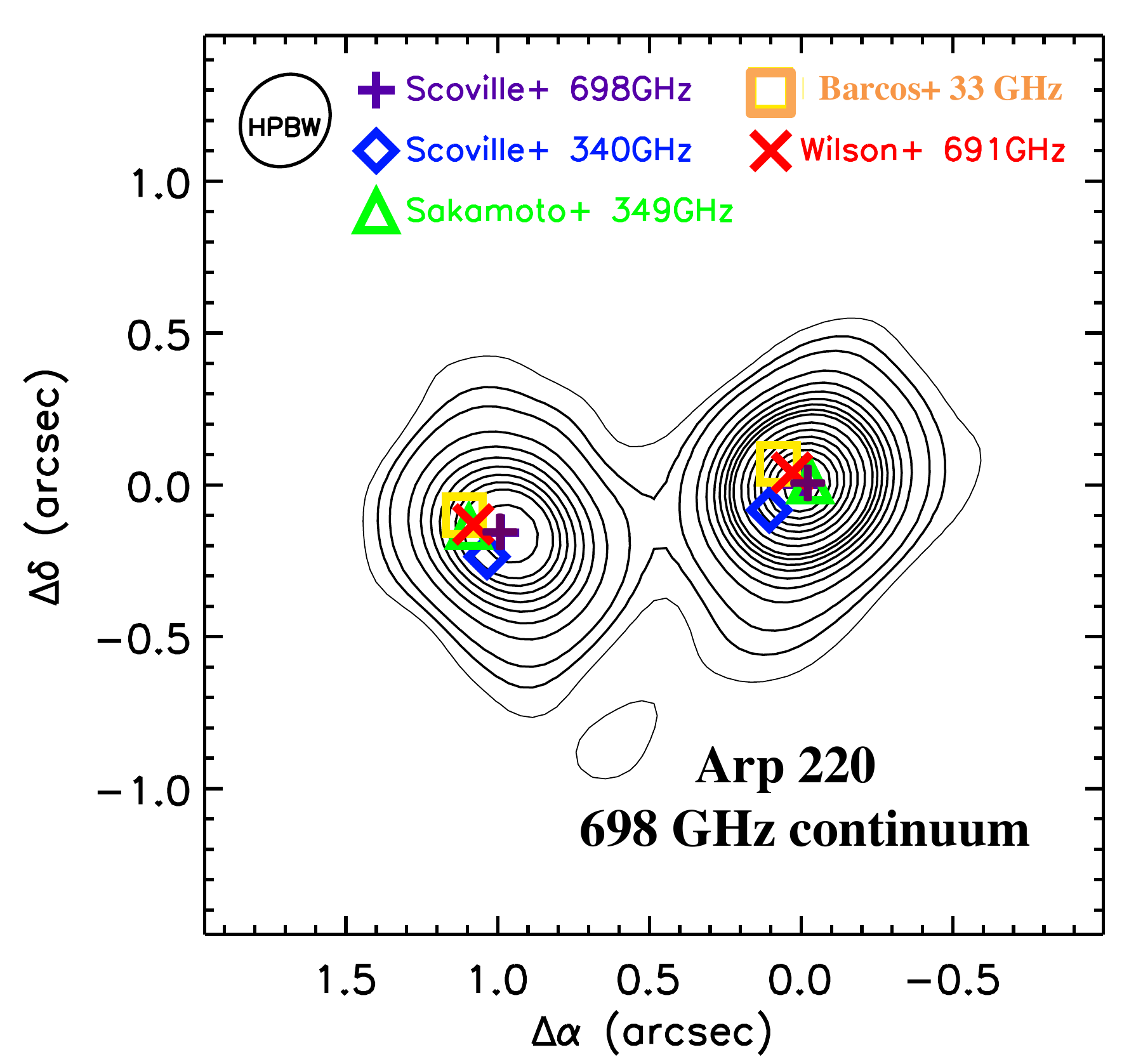}
\caption{The positions of the two nuclei in Arp 220 as derived from high resolution studies are compared: this work (Scoville \etal), 
\cite{sak08}, \cite{bar14} and \cite{wil14}. The contours represent our 
698 GHz continuum image (Figure \ref{continuum}). The position offsets are discussed in the text.} 
\label{arp_cont} 
\end{figure}

Both the delivered data products and our re-reduction of the Arp 220 datasets exhibited
non-flat residuals in line-free portions of the spectra after subtraction of the
continuum in the UV data. After attempting to use the much weaker
phase calibrator for passband calibration, we chose to use the
original passband calibrator and then did continuum subtraction in the 
image plane. This was done in IDL using our own
software which fit a polynomial to line-free frequency planes for each
spatial pixel, and then the endpoints of these polynomials were used to
linearly interpolate across the frequency planes with possible line
emission.


\section{Observational Results}\label{results}

In this section, we present the results for both sources, first the continuum (\S \ref{contin}) and then the line imaging (\S \ref{line}), followed 
by source flux measurements (\S \ref{measurements}). The source measurements are made by integration over circular apertures (Table \ref{apert}) and by fitting multiple Gaussian components (Table \ref{fits}). Synthesized beam parameters are given in Table \ref{sizes}. For Band 7, the typical beam size is 0.5\arcsec ~and for Band 9 they are 0.3\arcsec (see Table \ref{sizes}). 
For the imaging we use coordinate offsets from Arp 220 West, i.e. the Gaussian component fit to the 698 GHz continuum ($\alpha_{2000}=$15:34:57.217 and $\delta_{2000}=$23:30:11.44)
and for NGC 6240, the Gaussian-fit continuum peak at 694 GHz ($\alpha_{2000}=$16:52:58.898 and $\delta_{2000}=$2:24:03.51). 

\medskip 
\subsection{Continuum}\label{contin}

At submm wavelengths the continuum in ULIRGs
is dominated by dust emission, specifically the Rayleigh-Jeans (RJ) tail of the far-infrared dust emission. The total fluxes measured for Arp 220
and NGC 6240 are 490 and 20 mJy at 345 GHz and 4.51 and 0.126 Jy at 698 GHz  (Table \ref{apert}). The synchrotron and free-free emission which is responsible for the  longer wavelength radio continuum 
 is  $<30$ and $<16$ mJy at $\nu = 85$ GHz in Arp 220 and NGC 6240, respectively \citep{ima07,nak05}. And of course, both the synchrotron and free-free emissions decrease at higher frequency (varying as $\sim \nu^{-0.7}$ and $\nu^{-0.1}$ respectively). Their contributions to the fluxes observed here are therefore negligible. The submm-wavelength dust emission is generally optically thin and the observed RJ continuum fluxes can be used to  probe the ISM mass distributions \citep{sco14}. In \S \ref{dust_mass} we use the measured fluxes to estimate the ISM masses for each galaxy nucleus. 
 
The continuum images of Arp 220 and NGC 6240 at 345 GHz (Band 7) and 698 GHz (Band 9) are shown in Figure \ref{continuum}. The image in Figure \ref{continuum}-lower-left is similar in angular resolution and frequency to that recently obtained by \cite{wil14} with ALMA. The 342 GHz image (upper-left) has $2\times$ lower angular resolution but  much higher signal-to-noise ratio than the 350 GHz SMA map of  \cite{sak08}. 
 
Since the long wavelength dust emission is optically thin, the centroid of the dust emission may be used to pinpoint the ISM center of mass in the galaxy nucleus. In Figure \ref{arp_cont} we show the 698 GHz continuum contours together with the nuclear emission peaks
 found in previous high resolution imaging: the 33 GHz radio continuum \citep{bar14}, the 349 GHz continuum \citep{sak08}, and the 691 GHz continuum \citep{wil14}. 
 The peak positions obtained here (the plus and diamond symbols in Figure \ref{arp_cont}) were derived from the Gaussian-fit components as were those from \cite{sak08} and \cite{wil14}. The peak positions all agree to within 0.2\arcsec. The small shifts in the submm peaks, shown in Figure \ref{arp_cont}, could be caused by depth effects for the dust emission, or they may simply reflect uncertainties in the interferometric phase calibrations for Cycle 0 ALMA at high frequencies. In the following, we adopt our 698 GHz Gaussian-fit peaks (Table \ref{fits}) as the best positions for the 
 two galaxy nuclei.  
 
The total continuum fluxes at 350 GHz (870$\mu$m) and 700 GHz (430$\mu$m) are: 0.490 Jy and 4.51 Jy for Arp 220 (Table \ref{continuum}). Single dish (JCMT-SCUBA) fluxes for Arp 220 are: $832\pm86$ and $455\pm47$ mJy (average = 643 mJy) at 850$\mu$m \citep{dun00,ant04} and $6.3\pm0.8$ Jy at 450 $\mu$m \citep{dun01}.  Taking these fluxes at 'face value', our ALMA imaging is detecting approximately 76 and 71\% 
of the total flux at 350 and 700 GHz. 
For NGC 6240 we obtain integrated fluxes of 18-24 and 126 mJy (Table \ref{continuum}), compared with the JCMT-SCUBA total fluxes of 0.15 and 1 Jy \citep[$\pm30$\% --][]{kla01} at 850 and 450$\mu$m. Here, the detected percentages are only $\sim15$\%. In NGC 6240, the dust and line emission is clearly more extended than in Arp 220 so it is expected that the recovered flux percentage will be smaller than in Arp 220. However, the extremely small 'apparent' percentage (15\%) in NGC 6240 
is surprising since the telescope configuration should have good flux recovery for angular scales $<4$ \arcsec. It appears possible that there are significant calibration errors in the single dish measurements.  

\subsection{Lines}\label{line}

\begin{figure*}[t]
\plottwo{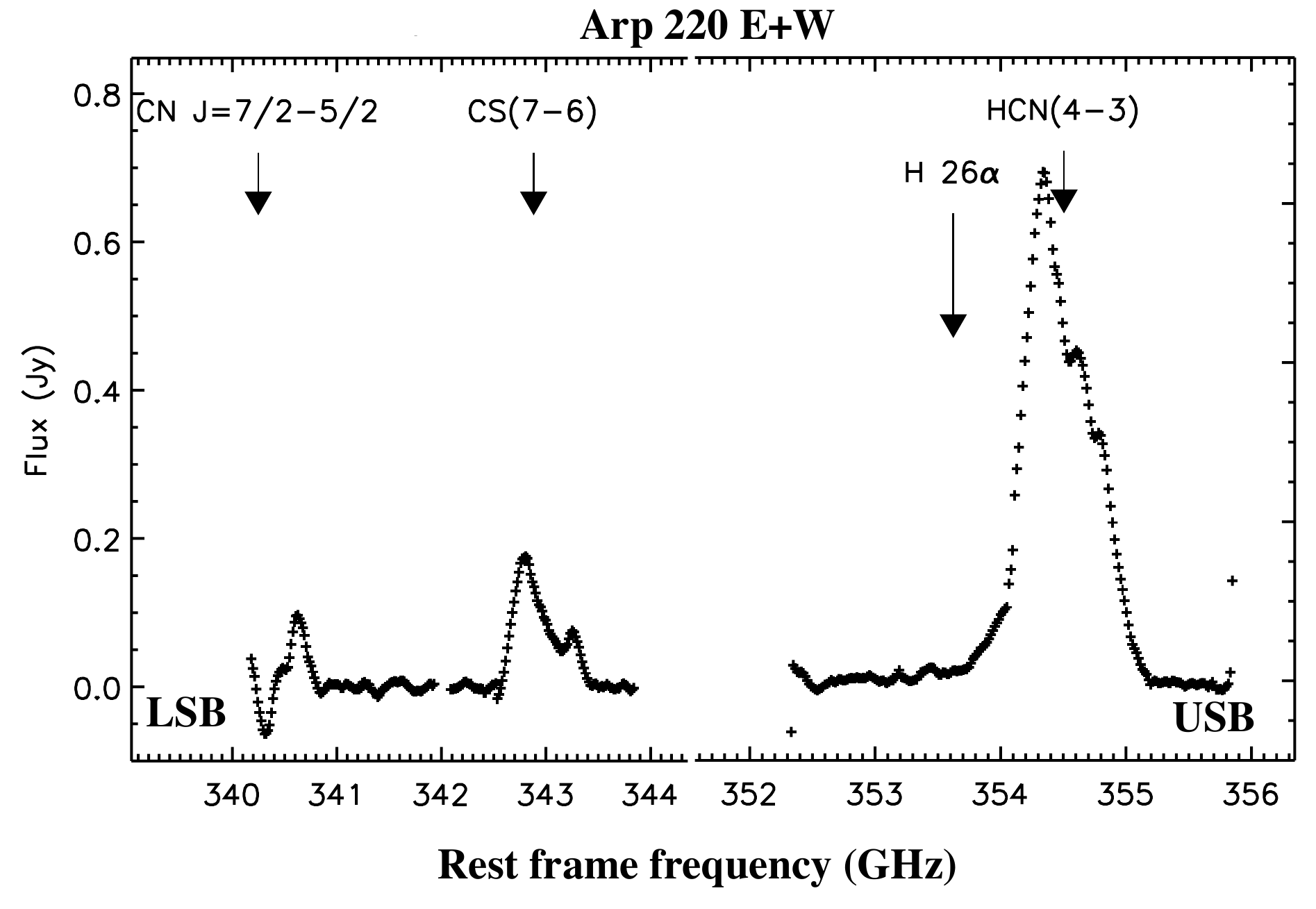}{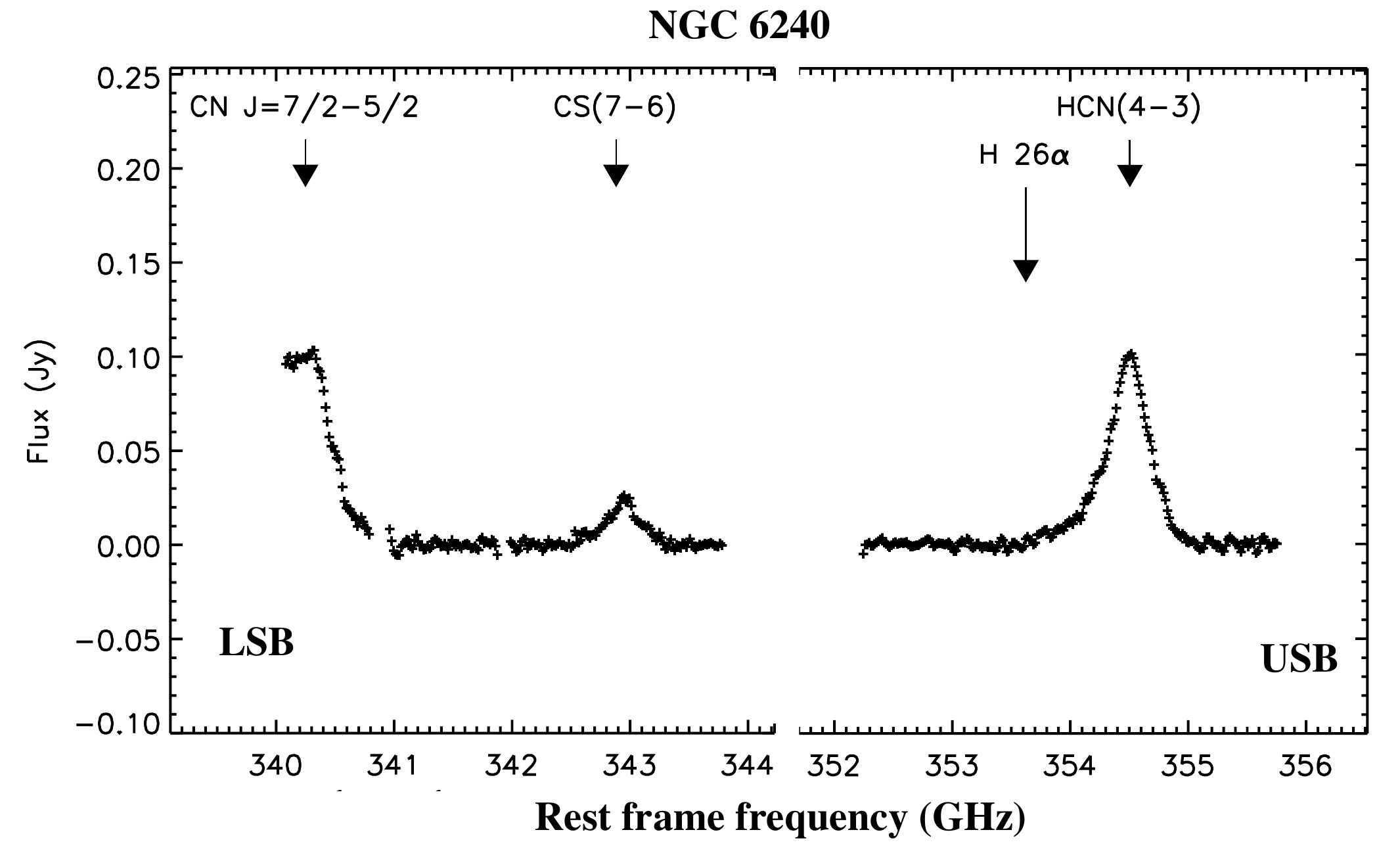}
\caption{Continuum-subtracted spectra of Arp 220 (left) and NGC 6240 (right) integrated over circular apertures of 1.8 and 3\arcsec ~diameter, respectively. The frequencies of the CN, CS 
and HCN lines are indicated by the arrows and the gap in frequency corresponds to the separation between the lower and upper sidebands (LSB \& USB).
The location of the H 26$\alpha$ recombination is indicated on the wing of the HCN emission. In view of the contamination by HCN, we defer discussion of the recombination lines \citep{sco13a} to our scheduled 
ALMA Cycle 2 data which includes other recombination lines.} 
\label{spect_total} 
\end{figure*}

\begin{figure*}[t]
\plottwo{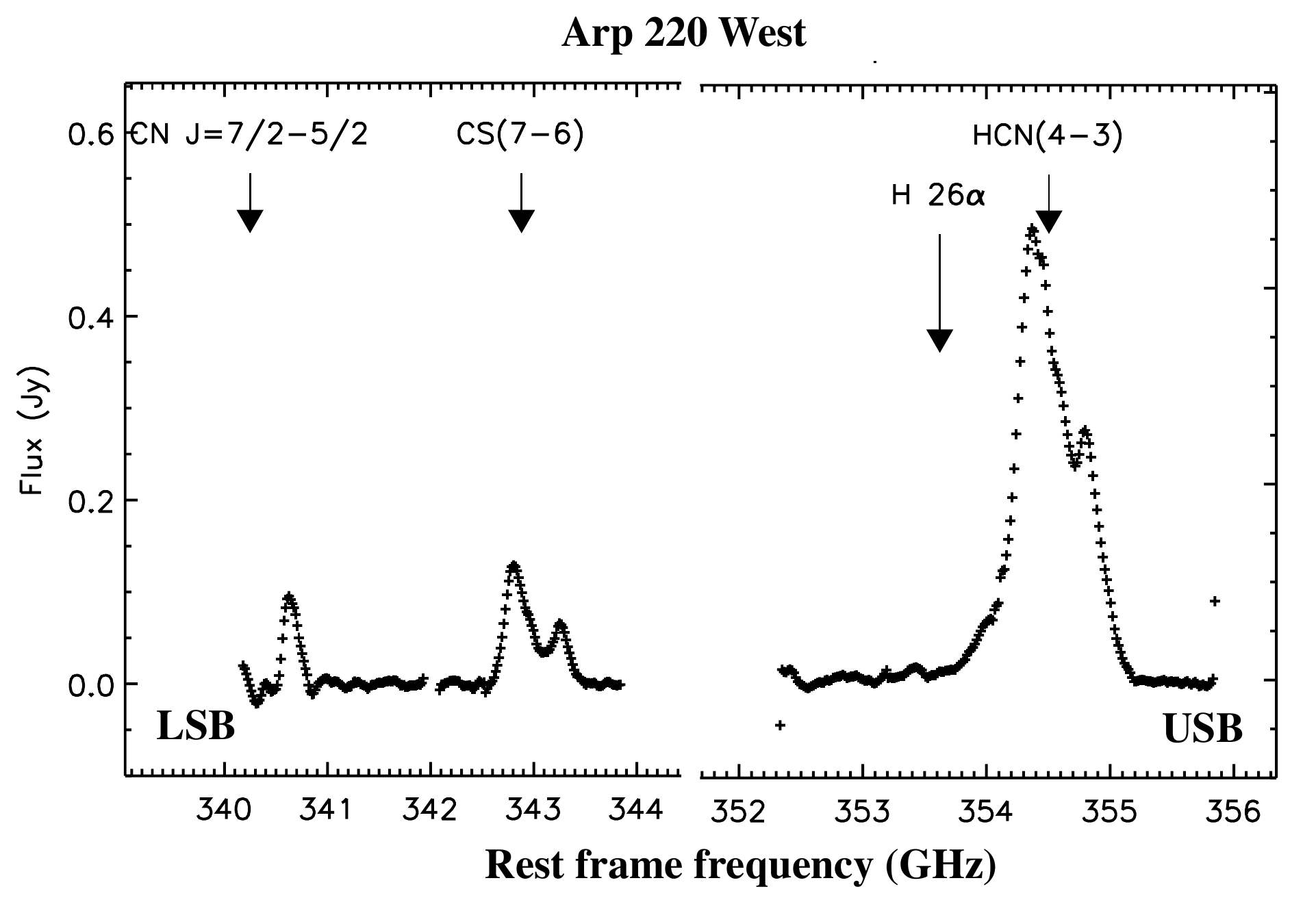}{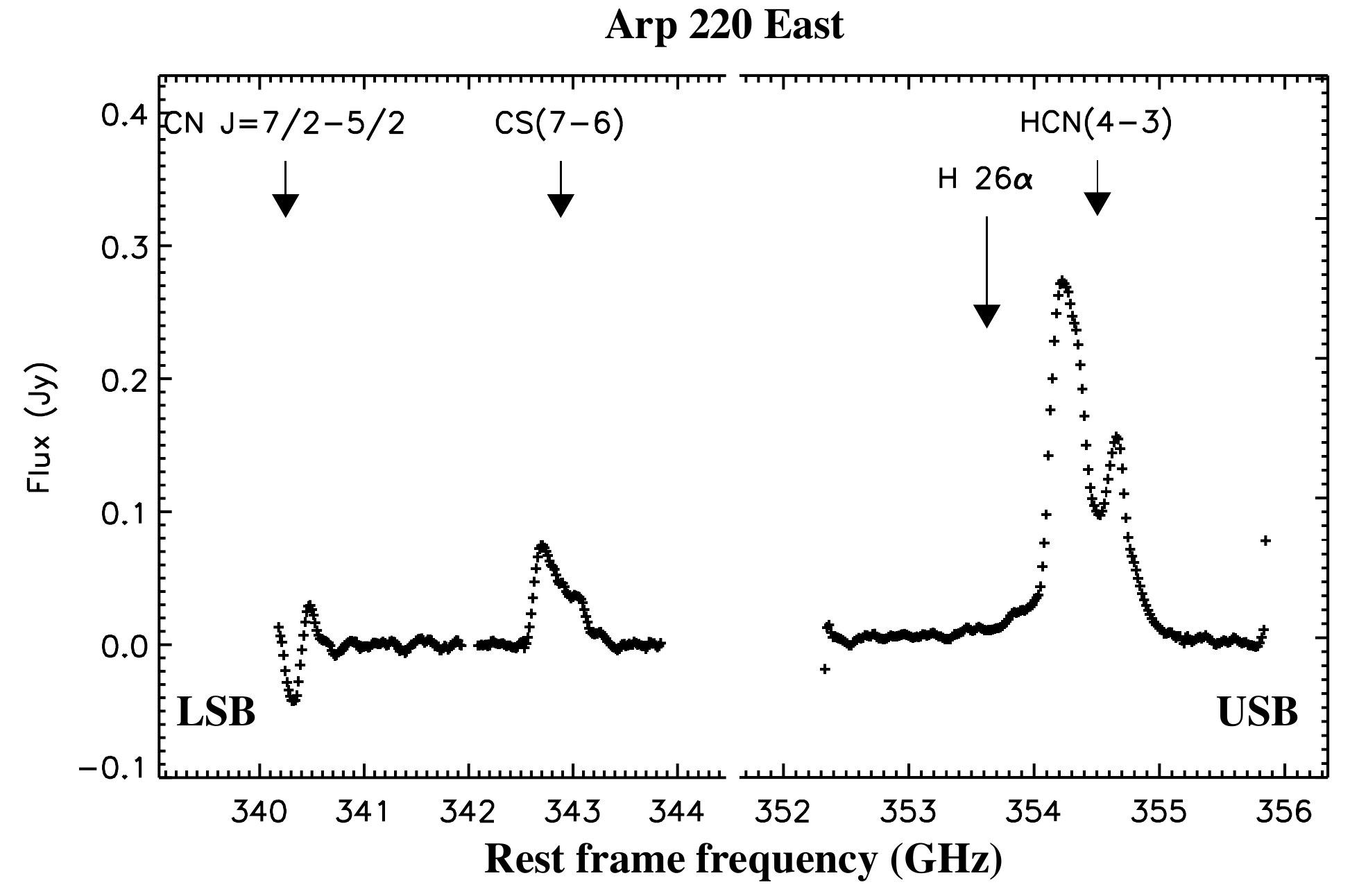}
\caption{Continuum-subtracted spectra of the Arp 220 nuclei integrated over circular apertures of 1.2\arcsec ~diameter centered on each peak. The location of the H 26$\alpha$ recombination is indicated on the wing of the HCN emission. In view of the contamination by HCN, we defer discussion of the recombination lines \citep{sco13a} to our scheduled 
ALMA Cycle 2 data which includes other recombination lines. 
} 
\label{spect_arp220} 
\end{figure*}

Figure \ref{spect_total} shows the source-integrated spectra for Arp 220 and NGC 6240 and Figure \ref{spect_arp220} shows the spectra for the individual East and West nuclei in Arp 220. The HCN (4 - 3) and CS (7 - 6) lines  are seen at very high signal-to-noise ratios in all of the spectra with a similar ratio HCN (4 - 3) / CS (7- 6) $\sim 0.2$ (see Table \ref{apert}). 
The CN (7/2 - 5/2) occurs at the low frequency end of the lower sideband (LSB) and thus the redshifted side of the emission 
is truncated. In both nuclei of Arp 220 this line appears in emission and absorption at different velocities (see Figure \ref{spect_arp220}). In NGC 6240 the CN emission is similar in strength to the HCN line (Figure \ref{spect_total}). 

The HCN and CS lines show double peaked velocity profiles in both nuclei of Arp 220 (Figure \ref{spect_arp220}). This is possibly 
due to absorption by optically thick, lower excitation gas at the systemic velocity along the line-of-sight to the nuclei. \cite{sak09} 
detect absorption in HCO$^+$ (3 - 2 \& 4 - 3) and CO (3 - 2) in small apertures on both nuclei. The deepest absorption in their 
spectra occurs at $5200 - 5450$ \kms, corresponding to the high frequency side of the HCN and CS line profiles where the dip is seen. The CN (7/2 - 5/2) line shows strong absorption on both nuclei of Arp 220 but, since this line is near the edge of the bandpass, we will not discuss it further. 


\begin{figure*}[t]
\plotone{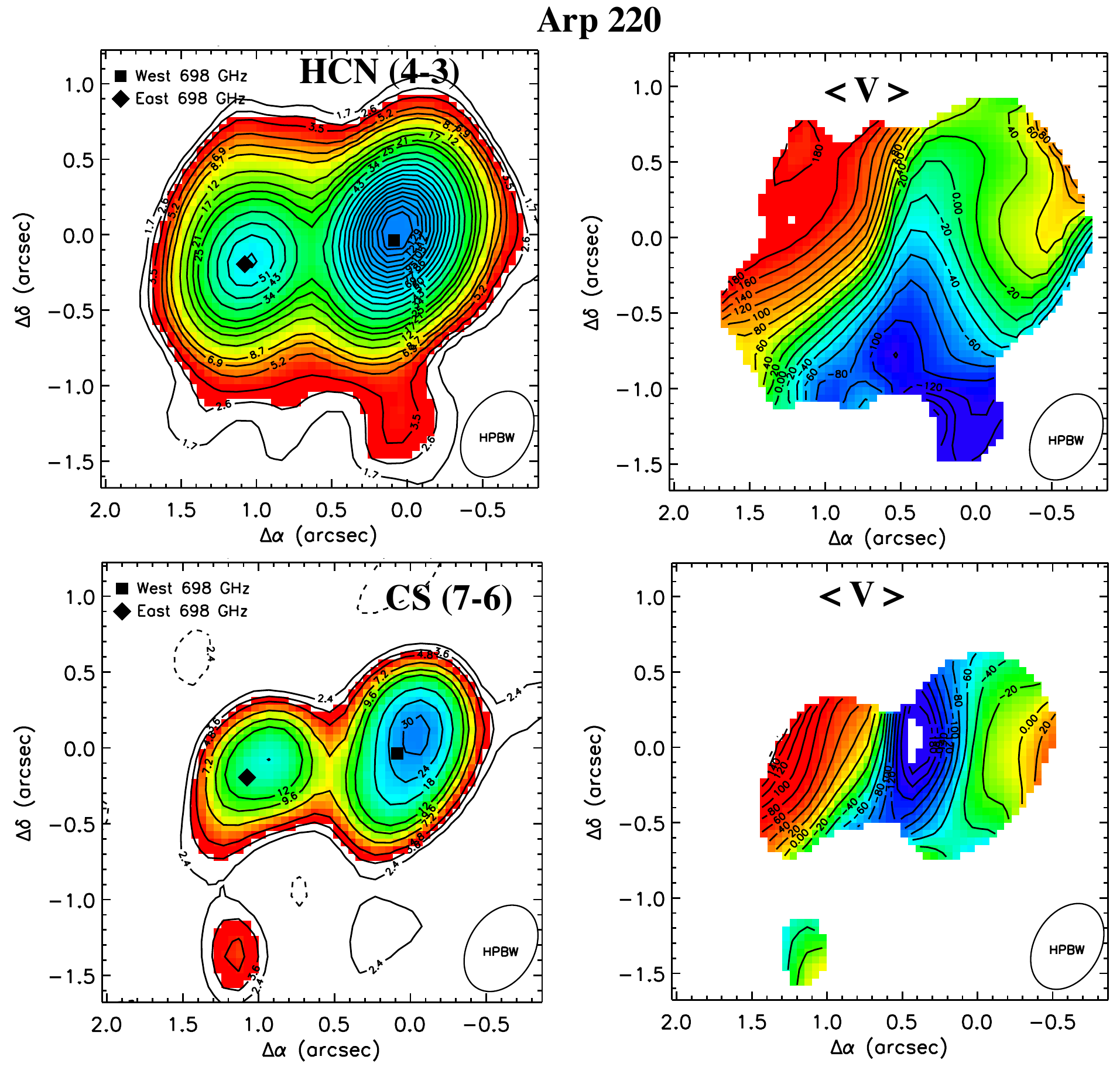}
\caption{Arp 220: Integrated line intensity images and mean velocities for the HCN (4 - 3) (top panels) and CS (7 - 6) emission (bottom panels). The square and diamond markers indicate the West and East continuum peaks at 698 GHz (see Table \ref{fits}) and the coordinate offsets 
are from the West continuum peak. Contours for 
HCN and CS are at -2 (dashed), 2, 3, 4, 6, 8, 10, 15, 20, 25, 30, 40, 50,... $ \times \sigma$
where $\sigma =  0.78$ and 1.2 Jy beam$^{-1}$ \kms. The velocity contour spacing is 20 \kms and the values are labelled.  The beam sizes ($0.52\times0.38$\arcsec) are shown.} 
\label{arp_line} 
\end{figure*}

\begin{figure*}[t]
\plotone{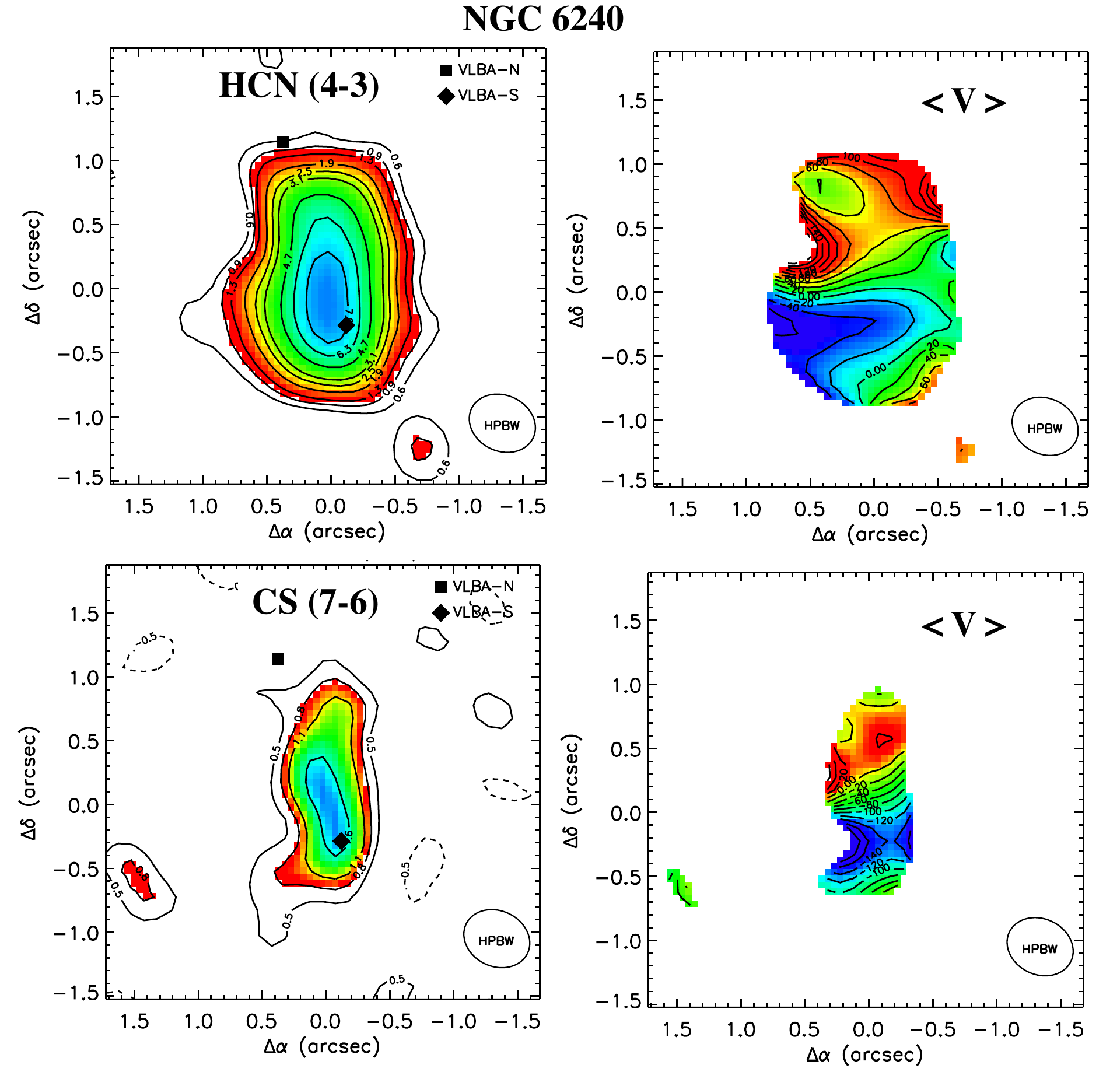}
\caption{NGC 6240: Integrated line intensity images and mean velocities for the HCN (4 - 3) (top panels) and CS (7 - 6) emission (bottom panels). Contours are labelled in Jy beam$^{-1}$ \kms and the levels for 
HCN and CS are at -2 (dashed), 2, 3, 4, 6, 8, 10, 15, 20 and 25 $ \times \sigma$
where $\sigma =  0.31$ and 0.25 Jy beam$^{-1}$ \kms. The velocity contour spacing is 20 \kms and the values are labelled.  The beam sizes ($0.55\times0.45$\arcsec) are shown.} 
\label{6240_line} 
\end{figure*}

Figures \ref{arp_line} and \ref{6240_line} show the velocity-integrated line fluxes and mean velocities for HCN and CS in Arp 220 and NGC 6240. Both the integrated intensity and the mean velocities are calculated 
as simple moment integrals over $V = \pm 500$\kms (without a signal-to-noise clipping threshold for the 3d pixels). The velocities 
are measured relative to the adopted systemic redshifts $z= 0.01812$ and $0.02448$. In both galaxies, the 
projected emissivity and velocity distributions are similar between HCN (4 - 3) and CS (7 - 6) -- the peaks coincide 
within 0.1\arcsec ~and the sizes are similar (see Table \ref{fits}). 

In Arp 220 West, 
the apparent kinematic major axes for HCN and CS are misaligned by $\sim45$\deg (PA = -45\deg for HCN and PA $\sim-90$\deg for CS, see Figure \ref{arp_line}). Most of this disagreement is on the western side of the West nucleus where the CS intensities are low. Lastly, there is also a minor difference along the southern 
boundary of the nuclei -- the East source showing a tongue of emission to the south in CS whereas in HCN a southerly extension is seen from the West nucleus (see Figure \ref{arp_line}). Some of these differences may be due to the very different optical depths of the CS (7-6) and HCN (4-3) emission lines (see Figure \ref{hcn_cs_excite}-Right). 

In NGC 6240 both the HCN and CS lines are much weaker than in Arp 220 so the signal-to-noise ratios (SNR) are not as high. 
In fact, the fitting of a single Gaussian component to the CS emission distribution did not converge due to low SNR (see Table \ref{fits}); however,  a fit was obtained  for HCN. Comparing the relative placement of the ISM  concentrations and the galaxy 
nuclei it is clear that in NGC 6240 the ISM is distributed along a ridge between the two nuclei \citep[as located 
by the VLBA N and S sources and the IR nuclei in][]{max07} rather than on the nuclei (as in Arp 220). This internuclear concentration of the molecular gas has been noted in previous investigations \citep{tac99,bry99,eng10}. However, the major peak of the submm dust emission and the HCN and CS is 
clearly nearer the South nucleus (see Figures \ref{continuum} and \ref{6240_line}). 

\begin{figure*}[t]
\plotone{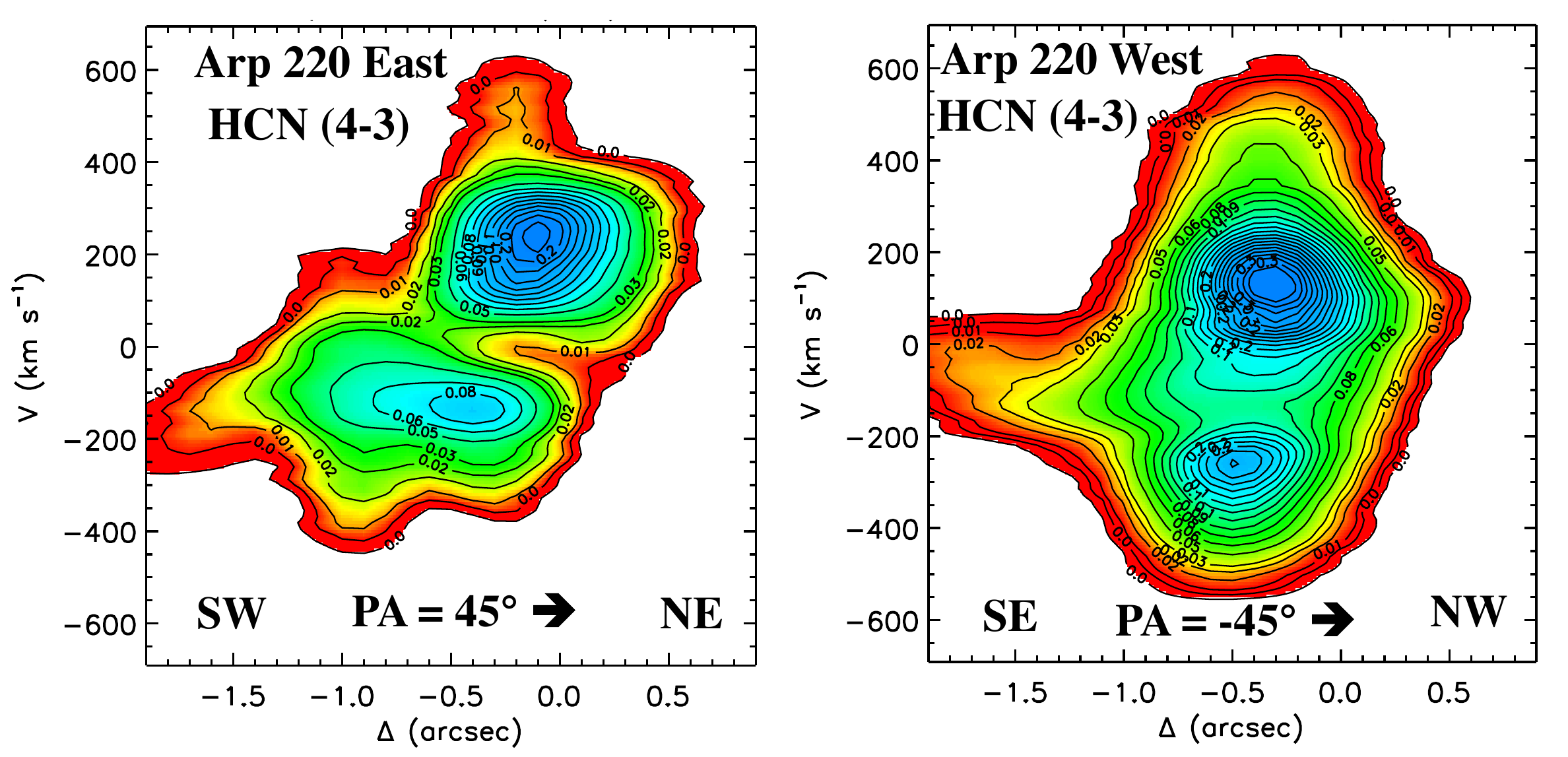}
\caption{Spatial-velocity strip maps in HCN (4 - 3) for Arp 220 East and West along their respective kinematic major axes. 
Velocity offsets of 200 and 400 \kms ~were subtracted at the systemic redshift z=0.01813.  Contours are at -2 (dashed), 2, 3, 4, 6, 8, 10, 15, 20, 25, ... $ \times \sigma$ where $\sigma =  3 $mJy beam$^{-1}$. } 
\label{arp_line_sv} 
\end{figure*}

\subsection{Measurements and Source Components}\label{measurements}

Aperture flux measurements for the continuum and the spectral lines 
are given in Table \ref{apert}. The table includes line fluxes 
integrated in velocity over $\pm500$\kms. 

We also computed least squares fits of elongated Gaussians to the observed brightness distributions and the fit parameters are given in Tables \ref{fits} and \ref{sizes}. 
For Arp 220 both nuclei were fit simultaneously with 2d Gaussians; the fits then provide a reasonable decomposition of the emission in the areas of overlap. In NGC 6240 
a double Gaussians were simultaneously fit to the peak in the south plus the extended tail to the north.  Table \ref{sizes}
includes an estimate of the true source sizes, with the beam analytically deconvolved. The peak brightness temperatures for both the observed and deconvolved 
 emission distributions are given in Tables \ref{fits} and \ref{sizes}, respectively.  For Arp 220, the peak continuum deconvolved brightness temperatures reach 
 10 - 40 K in Band 7 and 50 -150 K in Band 9. The black body temperature expected for the 
 measured source size and luminosity is $T_{bb} = 100$ K (\S \ref{dust_mass}). It is likely that the dust emission is optically thick in Band 9 but optically thin in 
 Band 7 \citep{dow07,mat09,wil14}. 
 
 The source sizes are probably best characterized by the dust continuum distribution using observations where the dust emission is optically thin 
 -- Band 7 for Arp 220 and Band 9 
 for NGC 6240. In Arp 220 West the deconvolved size is $0.36\times0.24$\arcsec (361 pc per \arcsec), implying a projected size of $130\times87$ pc. For Arp 220
 East, the size is $0.38\times0.32$\arcsec, or $137\times116$ pc. If these are modeled as inclined disks the implied inclinations are 48\deg and 31\deg from these axis ratios (the deconvolved sizes are uncertain given that they are less than the observational resolution).

\begin{deluxetable}{lclrcccccc}
\tabletypesize{}
\tablecaption{\bf{Aperture Measurements} \label{apert}  }
\tablewidth{0pt}
\tablehead{\colhead{Source} & \colhead{apert.\tablenotemark{a}} & \colhead{} & \colhead{total\tablenotemark{b}}  }  \\ 
\startdata
 & \multicolumn{3}{c}{\bf{lines~~~~~~~}}\\
\multicolumn{1}{c}{\bf{}}   & \arcsec & & Jy km s$^{-1}$ &  \\
\hline \\
Arp 220 total &   1.8   & HCN(4 - 3)    & 331$\pm$30  \\
Arp 220 East &    1.2 &      HCN(4 - 3)     & 106$\pm$11   \\
Arp 220 West &     1.2 &     HCN(4 - 3)     & 231 $\pm$23 \\
\\
Arp 220 total &   1.8 &    CS(7 - 6)      & 62$\pm$6  \\
Arp 220 East &   1.2 &    CS(7 - 6)      & 24$\pm$2   \\
Arp 220 West &   1.2 &    CS(7 - 6)    &  42$\pm$4  \\
\\
NGC 6240 &  3.0 & CN        & $>$39\tablenotemark{c}  \\
NGC 6240 &  3.0  & CS(7 - 6)          & 7.4 $\pm$1 \\
NGC 6240 &  3.0 & HCN(4 - 3)       & 36.9$\pm$4\\
\\
 & \multicolumn{3}{c}{\bf{continuum}}\\
\multicolumn{1}{c}{\bf{}}  & & GHz & Jy  ~~~~~~~& Jy beam$^{-1}$\\
\hline \\
 & \multicolumn{3}{c}{\bf{Band 7}}\\
Arp 220 total &    1.8 &  335.9   &        0.458$\pm$0.012 & 0.234$\pm$0.03\\
Arp 220 total &    1.8 &  347.6  &     0.490$\pm$0.002 & 0.288$\pm$0.01  \\
Arp 220 East &    1.2 &  335.9    &      0.152$\pm$0.017 & 0.106$\pm$0.04   \\
Arp 220 East &    1.2 &  347.6    &    0.161$\pm$0.017 & 0.119$\pm$0.02   \\
Arp 220 West &   1.2 &    335.9   &      0.313$\pm$0.007 & 0.234$\pm$0.03   \\
Arp 220 West &   1.2 &    347.6   &       0.342$\pm$0.002 & 0.288$\pm$0.01   \\
\\
NGC 6240 & 3.0 & 333.9    &       0.024$\pm$0.002 & 0.008$\pm$0.01  \\
NGC 6240 & 3.0 & 345.3    &     0.018$\pm$0.002 & 0.006$\pm$0.01  \\
    \\ 
 & \multicolumn{3}{c}{\bf{Band 9}}\\
Arp 220 total &    1.8 &  697.7   &        4.14$\pm$0.07 & 2.75$\pm$0.02\\
Arp 220 East &    0.8 &  697.7    &      2.15$\pm$0.15 & 1.36$\pm$0.06   \\
Arp 220 West &   0.8 &   697.7   &      3.80$\pm$0.04 & 2.75$\pm$0.03   \\
\\
NGC 6240 & 3.0 & 693.5    &       0.106$\pm$0.004 & 0.025$\pm$0.01  \\

\enddata
\tablecomments{Total integrated fluxes were measured in circular apertures of diameter given in column 2. The peak fluxes were 
measured with the synthesized beams given in Table \ref{sizes} but restricted to the same diameter regions.}
\tablenotetext{a}{Measurement aperture diameter in arcsec.}
\tablenotetext{b}{Uncertainties on the line fluxes are taken to be the the larger of  the 10\% uncertainty in flux calibration or the measured rms noise. The uncertainties
for the continuum measurements were derived from the dispersion in measurements taken in equal size apertures centered at
off-source positions.}
\tablenotetext{c}{Lower limit since the line is on the edge of the bandpass.}
\end{deluxetable}

\begin{deluxetable*}{llrrrrr}
\tabletypesize{\scriptsize}
\tablecaption{\bf{Gaussian Fit Components} \label{fits}  }
\tablewidth{0pt}
\tablehead{\colhead{Source} & \colhead{line/continuum} & \colhead{$\nu_{obs}$} & \colhead{$\alpha_{2000}$} & \colhead{$\delta_{2000}$} & \colhead{peak Flux}  & \colhead{peak $T_B$}    \\
\colhead{}  & \colhead{}  &  \colhead{GHz} &  \colhead{} & \colhead{} &  \colhead{} & \colhead{K} }
\startdata
\\ 
 &&\multicolumn{3}{c}{\bf{Band 7}}\\
\\     & & & & & Jy beam$^{-1}$ \\
     Arp 220 W &  continuum &  335.86 &  15:34:57.224 &  23:30:11.357 &    0.238 $\pm$ 0.0007 &  10.2 \\ 
     Arp 220 E &  continuum &  335.86 &  15:34:57.286 &  23:30:11.206 &    0.105 $\pm$ 0.0007 &   4.5 \\ 
     Arp 220 W &  continuum &  347.60 &  15:34:57.224 &  23:30:11.357 &    0.293 $\pm$ 0.0016 &  14.6 \\ 
     Arp 220 E &  continuum &  347.60 &  15:34:57.286 &  23:30:11.199 &    0.118 $\pm$ 0.0015 &   5.9 \\ 
    NGC 6240 S &  continuum &  333.87 &  16:52:58.909 &    2:24:03.203 &    0.007 $\pm$ 0.0003 &   0.3 \\ 
    NGC 6240 N &  continuum &  333.87 &  16:52:58.916 &    2:24:04.177 &    0.003 $\pm$ 0.0002 &   0.1 \\ 
    NGC 6240 S &  continuum &  345.35 &  16:52:58.909 &    2:24:03.200 &    0.005 $\pm$ 0.0001 &   0.2 \\ 
    NGC 6240 N &  continuum &  345.35 &  16:52:58.920 &    2:24:04.232 &    0.002 $\pm$ 0.0001 &   0.1 \\ 
    \\ 
 && \multicolumn{3}{c}{\bf{Band 9}}\\
\\
     Arp 220 W &  continuum &  697.69 &  15:34:57.217 &  23:30:11.440 &    2.672 $\pm$ 0.0053 &  75.2 \\ 
     Arp 220 E &  continuum &  697.69 &  15:34:57.283 &  23:30:11.282 &    1.316 $\pm$ 0.0048 &  37.0 \\      
              NGC 6240 &  continuum &  693.46 &  16:52:58.898 &    2:24:3.512 &    0.022 $\pm$ 0.0001 &   0.9 \\ 
              \\
          && \multicolumn{3}{c}{\bf{Lines}}\\

    & & & & & Jy km s$^{-1}$ beam$^{-1}$ & \tablenotemark{b} \\
       Arp220 W & CS (7 - 6) &  336.71 &  15:34:57.219 &  23:30:11.400 &   24.730 $\pm$ 0.0003 &   4.73 \\ 
        Arp220 E & CS (7 - 6) &  336.71 &  15:34:57.287 &  23:30:11.221 &   15.636 $\pm$ 0.0003 &   2.86 \\ 
       Arp220 W & HCN (4 - 3) &  348.18 &  15:34:57.222 &  23:30:11.383 &  143.213 $\pm$ 0.0005 &   17.20 \\ 
       Arp220 E & HCN (4 - 3) &  348.18 &  15:34:57.288 &  23:30:11.214 &   56.775 $\pm$ 0.0005 &   10.30 \\ 
      NGC 6240 & CS (7 - 6) &  334.65 &  ....\tablenotemark{a} &   ....\tablenotemark{a} & ...\tablenotemark{a} &  0.27  \\ 
     NGC 6240 & HCN (4 - 3) &  345.99 &  16:52:58.907 &    2:24:03.559 &    8.985 $\pm$ 0.0001 &   1.10 \\ 
  \enddata
\tablecomments{Peak position and fluxes obtained from two component 2d gaussian fits. Gaussian component sizes are listed in Table \ref{sizes}. Uncertainties in the continuum and line fluxes obtained from the uncertainties in the Gaussian component fitting do not include 
calibration uncertainties.}
\tablenotetext{ a}{No convergent fit obtained.}
\tablenotetext{ b}{Peak brightness temperature from the spectral cube within $\pm 500$ \kms ~of the systemic velocity.}
\end{deluxetable*}

\begin{deluxetable*}{llllrrrcrrrrrr}
\tabletypesize{\scriptsize}

\tablecaption{\bf{Gaussian Source Sizes from Table \ref{fits}} \label{sizes}  }
\tablewidth{0pt}
\tablehead{  &  &\multicolumn{3}{c}{\bf{Beam}} &\multicolumn{3}{c}{\bf{Source}} &\multicolumn{3}{c}{\bf{Deconvolved}}\\
  &  & \multicolumn{3}{c}{--------------------------------} & \multicolumn{3}{c}{------------------------------------------------------------} & \multicolumn{3}{c}{----------------------------}\\
\colhead{Source} & \colhead{} &  \colhead{major} & \colhead{minor} &  \colhead{PA}  &  \colhead{major} & \colhead{minor} &  \colhead{PA}  &  \colhead{major} & \colhead{minor} &  \colhead{T$_B$}   \\
\colhead{}  & \colhead{}  & \colhead{\arcsec} &  \colhead{\arcsec} &\colhead{\deg} & \colhead{\arcsec} &  \colhead{\arcsec} &\colhead{\deg}  & \colhead{\arcsec} &  \colhead{\arcsec} &\colhead{K}}
\startdata
\\ 
 &&&&& \multicolumn{3}{c}{\bf{Band 7}}\\
\\
    Arp 220 W &  continuum &  0.60 &  0.42 &  -32.0 &   0.66 $\pm$ 0.01 &   0.50 $\pm$ 0.01 &  -29 $\pm$     1 &  0.28 &  0.27 &  33.9 \\ 
     Arp 220 E &  continuum &  0.60 &  0.42 &  -32.0 &   0.71 $\pm$ 0.02 &   0.50 $\pm$ 0.01 &  -16 $\pm$     1 &  0.37 &  0.27 &  11.1 \\ 
     Arp 220 W &  continuum &  0.52 &  0.39 &  -27.2 &   0.63 $\pm$ 0.01 &   0.46 $\pm$ 0.02 &  -28 $\pm$     1&  0.36 &  0.24 &  34.9 \\ 
     Arp 220 E &  continuum &  0.52 &  0.39 &  -27.2 &   0.65 $\pm$ 0.01 &   0.51 $\pm$ 0.01 &  -17 $\pm$     3 &  0.38 &  0.32 &   9.6 \\ 
    NGC 6240 S &  continuum &  0.55 &  0.46 &   65.6 &   0.74 $\pm$ 0.04 &   0.60 $\pm$ 0.03 & -152 $\pm$    13 &  0.50 &  0.39 &   0.4 \\ 
    NGC 6240 N &  continuum &  0.55 &  0.46 &   65.6 &   1.22 $\pm$ 0.11 &   0.70 $\pm$ 0.06 & -150 $\pm$     8 &  1.09 &  0.53 &   0.1 \\ 
    NGC 6240 S &  continuum &  0.53 &  0.44 &   64.7 &   0.72 $\pm$ 0.02 &   0.59 $\pm$ 0.02 & -155$\pm$     2 &  0.49 &  0.40 &   0.3 \\ 
    NGC 6240 N &  continuum &  0.53 &  0.44 &   64.7 &   1.19 $\pm$ 0.01 &   0.72 $\pm$ 0.02 & -149 $\pm$     1 &  1.07 &  0.57 &   0.03 \\ 
\\      
          &&&&& \multicolumn{3}{c}{\bf{Band 9}}\\
\\
     Arp 220 W &  continuum &  0.32 &  0.28 &  -38.6 &   0.39 $\pm$ 0.01 &   0.34 $\pm$ 0.01 &  -42 $\pm$     1 &  0.23 &  0.19 & 148.9 \\ 
     Arp 220 E &  continuum &  0.32 &  0.28 &  -38.6 &   0.44 $\pm$ 0.01 &   0.37 $\pm$ 0.02 & -139 $\pm$     1 &  0.30 &  0.24 &  47.2 \\ 
    NGC 6240 &  continuum &  0.27 &  0.24 &   29.7 &   0.86 $\pm$ 0.02 &   0.39 $\pm$ 0.02 & -173 $\pm$     1 &  0.82 &  0.30 &   0.2 \\ 
 \\    
          &&&&& \multicolumn{3}{c}{\bf{Lines}}\\
\\
        Arp220 W & CS (7 - 6) &  0.60 &  0.42 &  -32.0 &   0.78 $\pm$ 0.02 &   0.60 $\pm$ 0.02 &  -31 $\pm$     3 &  0.49 &  0.43 &   10.1 \\ 
        Arp220 E & CS (7 - 6) &  0.60 &  0.42 &  -32.0 &   0.72 $\pm$ 0.02 &   0.54 $\pm$ 0.02 &  -35 $\pm$     2 &  0.40 &  0.35 &   7.5 \\ 
       Arp220 W & HCN (4 - 3) &  0.52 &  0.39 &  -27.2 &   0.77 $\pm$ 0.01 &   0.57 $\pm$ 0.01 &  -24 $\pm$     1 &  0.57 &  0.41 &   39.3 \\ 
       Arp220 E & HCN (4 - 3) &  0.52 &  0.39 &  -27.2 &   0.78 $\pm$ 0.01 &   0.59 $\pm$ 0.01 & 20 $\pm$     2 &  0.58 &  0.45 &   21.5 \\ 
      NGC 6240 & CS (7 - 6) &  0.55 &  0.46 &  65.6 &  ....\tablenotemark{a} &   ....\tablenotemark{a} &   ....\tablenotemark{a} &  ....\tablenotemark{a} &   ....\tablenotemark{a} &    ....\tablenotemark{a} \\ 
     NGC 6240 & HCN (4 - 3) &  0.53 &  0.44 &   64.7 &   1.26 $\pm$ 0.02 &   0.74 $\pm$ 0.02 & -178 $\pm$     3 &  1.14 &  0.60 &   1.5 \\ 
   \\
\enddata
\tablecomments{Sizes and major axis PA estimates obtained from 2d gaussian fits listed in Table \ref{fits}. The uncertainties in the parameters were the formal errors from the multi-component Gaussian fitting.}
\tablenotetext{ a}{No convergent fit obtained.}
\end{deluxetable*}

\section{Long Wavelength Dust Continuum as a Mass Tracer}\label{dust_mass}

In this section we use the long wavelength dust emission to estimate the masses of dust and gas in each nucleus. Assuming the fluxes of the nuclei are well-recovered in the interferometry, this is
a very robust mass estimator. At sufficiently long wavelengths the dust is optically thin so the observations sample the entire 
column of dust and ISM. The emitted flux on the Rayleigh-Jeans tail of the spectrum is only linearly dependent on the dust temperature (T$_d$)
and for normal galaxies the bulk of the ISM mass is at typically 20--40K. Averaging over whole galaxies or galactic nuclei, the dust opacity coefficient and dust to gas mass ratio 
are also probably not strongly varying (at least for massive galaxies with near solar metallicity). Thus the long wavelength dust continuum 
can be used as a linear probe of the ISM mass, avoiding the problems associated with optically thick molecular lines with sub-thermal 
excitation. In application of this technique to galactic nuclei, one does need to allow for increased dust temperatures in the stronger radiation field of the nuclear regions The use of the dust continuum is fully developed along with empirical calibration of the flux to mass ratio in \cite{sco14} and is summarized briefly below.

At long wavelengths, the dust emission is optically thin and the observed flux density is given by 
\begin{equation}
S_{\nu} \propto  \kappa_{d}(\nu)  T_{\rm d} \nu^2 {M_{\rm d}\over{d_L^2}}
\label{fnu}
\end{equation}

\noindent where $T_{\rm d}$ is the temperature of the grains,  $\kappa_{d}(\nu)$ is the dust opacity per unit mass of dust, $M_{\rm d}$ is the total
mass of dust and $d_L$ is the luminosity distance. (For simplicity in illustrating the physics, Equation~\ref{fnu} does not include the effects of bandshifting and compression of frequency space which occur for high redshift sources.) In local star-forming galaxies, most of the mass of dust is at $T_d \sim 20 \rightarrow 30$ K \citep[see references in][]{sco14} and variations 
in the \emph {mass-weighted} dust temperatures between galaxies are small ($\sim$20 - 40K); the observed fluxes then probe the total mass of dust. 

To obviate the need to know both the dust opacity and dust-to-gas ratio (which are degenerate when using Equation \ref{fnu} to estimate ISM masses), one can instead empirically calibrate  
the ratio of the specific luminosity at rest frame 850$\mu$m to total ISM mass using samples of observed galaxies, thus absorbing the opacity curve and the abundance ratio into a single empirical constant $\alpha_{850\mu \rm m} =  L_{\nu_{850\mu \rm m}} / M_{\rm ISM} $.
In \cite{sco14}, three samples were developed: 1) 12 local star forming and starbursting galaxies with global SCUBA and ISM measures; 
2) extensive Galactic observations from the Planck Collaboration and 3) a sample of 28 SMGs at z $< 3$ having CO(1-0) measurements. The 3 samples yielded  $\alpha_{850\mu \rm m} $= 1.0, 0.79 and 1.01
$\times10^{20}  \rm ~ergs~ sec^{-1} Hz^{-1} \msun^{-1}$, respectively \citep[][]{sco14}. The Planck measurements are particularly convincing --  they are of high SNR, span a large wavelength range and they probe the diversity of Galactic ISM
including both HI and H$_2$ dominated clouds. The Planck measurements exhibit little variation in their empirical coefficient $\alpha_{250\mu \rm m}$ which is equivalent to our $\alpha_{850\mu \rm m}$ \citep{pla11a}. The Planck measurements also determine very well the long wavelength dust emissivity index: $\beta = 1.8 \pm 0.1$ \citep{pla11b}. Lastly, we note that \cite{dra07b} see no variation in the 
dust abundance down to $\sim20$\% of solar metallicity and it is reassuring that the low redshift galaxy sample and the high z SMGs yield very similar $\alpha_{850\mu \rm m}$.

Taking account of the frequency shifting of the rest frame IR spectral energy distribution as a function of redshift (z), and assuming one is 
observing the optically thin RJ tail with opacity index $\beta = 1.8$, \cite{sco14} show that the expected continuum flux is 

\noindent 
\begin{eqnarray}
S_{\nu} &=& 1.15 ~  {M_{\rm ISM} \over 10^{10} \msun} ~ {(1+z)^{4.8} ~T_{25}  ~ \nu_{350}^{3.8}  ~\Gamma_{RJ}~\over{  d_{Gpc}^{2}}}  \rm ~mJy  \nonumber \\
 \end{eqnarray}
 \noindent and the implied mass is therefore, 
 \begin{eqnarray}
M_{\rm ISM} = { 0.868 \times S_{\nu}(mJy)    d_{Gpc}^{2}          \over{ (1+z)^{4.8} ~T_{25}  ~ \nu_{350}^{3.8}  ~\Gamma_{RJ}~ }  }  \rm ~10^{10} \msun \nonumber \\
 \end{eqnarray}\label{ism_mass}

\noindent where the observed frequency is normalized to 350 GHz, the dust temperature is normalized to 25K and the luminosity distance to 1Gpc. 

$\Gamma_{RJ}$ is the correction factor for departure from the RJ $\nu^2$ dependence given by :
\begin{eqnarray}
\Gamma_{RJ}(T,\nu, z) &=&  0.672\times {\nu_{350} (1+z)      \over{ T_{25}}}  \times \nonumber \\
&& {1 \over{e^{0.672\times \nu_{350} (1+z) / T_{25}} -1 }}  ~~~~.
  \end{eqnarray}  \label{flux}

In the case of ULIRGs, there are two cautions with respect to application of Equation \ref{ism_mass}: 1) the observed wavelength must be sufficiently long that the 
dust emission is optically thin and 2) the mean dust temperature (mass-weighted) for the ISM within the compact nuclear region is likely to be elevated well above the typical global 
average of $\sim$25 K for ISM in normal galaxy disks. Modeling the IR SED of Arp 220, \cite{wil14} found that the dust optical depth was 
$\tau \sim 1.7$ and 5.3  at 678 GHz and $T_d = 80$ and 197 K for the East and West nuclei, respectively. Similarly, \cite{mat09} used earlier SMA observations at 690 GHz to conclude that the dust emission had $\tau \sim 1$ at 450 $\mu$m and  $T_d = 49$ and 97 K for the East and West nuclei. 

One could correct for modest optical depths by including 
a factor $f(\tau) = (1-e^{-\tau})/\tau$ in Equation \ref{ism_mass} but a more reliable 
mass estimate is obtained using the lower frequency continuum measurements where the opacity is much less. 
For the Arp 220 nuclei we therefore use the 350 GHz fluxes and adopt T$_d = 100$ K. This is recognized as an approximation since the 
two nuclei are different and undoubtedly have temperature gradients. A good derivation of the dust temperature from the observations 
would require multiple spectral bands, observed at the same resolution and with resolution sufficient to resolve likely temperature gradients. 

The adopted dust temperature is consistent with the luminosity outputs of the nuclei.  The Arp 220 
nuclei have $L_{FIR} \simeq 10^{12}$ \lsun; the blackbody radius required to emit this luminosity is 70 pc if the temperature is 100 K. This radius 
is comparable with the measured sizes of the dust emission peaks, implying that the dust must be heated to $\gtrsim 100$ K throughout this volume in order to emit the observed 
luminosity. In NGC 6240, the submm flux is
much lower and the emission region is much larger;  we therefore assume that the continuum is optically thin at 693 GHz and adopt the more typical mean T$_d \sim 25$ K, since the dust is more 
spatially extended and the IR luminosity is $\sim3 - 4$ times lower. 

Using 
these values in Equation \ref{ism_mass}, we find ISM mass estimates for the nuclei in Arp 220 (z = 0.018) and NGC 6240 (z = 0.024) in the range $1.6 - 6.0 \times 10^9$ \msun(see Table \ref{masses}). Given the uncertainties in the adopted mass-weighted dust temperature we expect these estimates to be uncertain by a factor of at most 2 for the nuclear sources. 

The mass estimates derived from the dust continuum are similar to those obtained from the CO (1- 0 and 2-1) emission lines. The total for both nuclei of Arp 220 is  $\sim5\times10^9$ \msun \citep{sco97}; the individual East and West nuclei are $\sim2-3\times10^9$ \msun \citep{dow98} and for NGC 6240 $3\times10^9$ \msun \citep{tac99}. 
The RJ dust continuum obviously avoids the uncertainties in the conversion factor (and the uncertain relative brightness of high J CO 
transitions). One issue which could arise in very dense ISM environments is grain growth by coagulation or grain destruction by collisions; these would change 
the long wavelength dust opacity coefficient in different directions. We can hope that future ALMA submm imaging at higher and matched resolutions 
at different wavelengths might demonstrate if the opacity spectral index is non-standard. 


Combining the masses derived from the RJ dust emission with the source radius determined from the deconvolved dust emission distributions (Table \ref{masses}), we obtain mean column densities of 
$<N_{H2}> \gtrsim$ 6, 15 and 0.7 $\times 10^{24}$ cm$^{-2}$ for Arp 220 East, West and NGC 6240, assuming a uniform, spherical distribution of ISM (see Table \ref{masses}). Estimates of $<N_{H2}> \sim 4\times 10^{25}$ cm$^{-2}$ were obtained by \cite{sak08}
for the Arp 220 nuclei, adopting $\tau_{850\mu m} \sim 1$ and using the \cite{hil83} Galactic dust opacity law. These are lower limits for the column density along the line of sight to the nuclei since they have assumed a uniform distribution of gas and it is more likely that the ISM is centrally peaked. The column density estimates obtained here are more constraining than those by \cite{sak08} due to the more up-to-date empirical value for the ISM mass to dust opacity coefficient. Since a gas column of $10^{25}$ cm$^{-2}$ would heavily absorb the soft X-rays, 
these column densities would obscure any X-ray emission from a central AGN in Arp 220 if such exists \citep{pag14}. Thus the non-detection of X-rays does not imply there is no AGN.

\begin{deluxetable*}{lrrrrccccc}
\tabletypesize{}
\tablecaption{\bf{ISM Masses from Dust Continuum}} 
\tablewidth{0pt}
\tablehead{\colhead{Source}  & \colhead{$\nu_{obs}$} & \colhead{Flux}   & \colhead{T$_d$ \tablenotemark{a}} & \colhead{$\Gamma_{RJ}$\tablenotemark{d}}   & \colhead{Mass}  & \colhead{diameter\tablenotemark{b}} &  \colhead{Radius} & \colhead{$<\Sigma_{gas}>$\tablenotemark{c}}  & \colhead{$<N_{H2}>$\tablenotemark{c}} \\
\colhead{}    & \colhead{} &  \colhead{} &  \colhead{} &  \colhead{} &  & & &  }
\startdata
&    GHz & mJy & K & & $10^9$~\msun & \arcsec & pc & $\msun ~\rm pc^{-2}$  & $ \rm cm^{-2}$ \\
\hline \\
Arp 220 total &      347.6  &     490 & 100 & 0.917 & 5.9  \\
Arp 220 East &      347.6    &    161&   100  & 0.917 & 1.9 & $\lesssim 0.38$ & $\lesssim 69$ & $\gtrsim 1.3\times10^5$ & $\gtrsim 6.0\times10^{24}$ \\
Arp 220 West &       347.6   &    342 &   100 & 0.917 & 4.2 & $\lesssim 0.36$ & $\lesssim 65$ & $\gtrsim 3.1\times10^5$ & $\gtrsim 1.5\times10^{25}$ \\
\\
NGC 6240 &  693.5    &   126 &  25 & 0.468 &  1.6 & 0.8 & 190 & $\gtrsim 1.4\times10^4$ & $\gtrsim 7.0\times10^{23}$\\
\enddata
\tablecomments{ISM masses derived from the RJ continuum flux using Equation \ref{ism_mass}
with distances of 74 and 103 Mpc for Arp 220 and NGC 6240.}
\tablenotetext{a}{Adopted dust temperature used for mass calculation as discussed in text.}
\tablenotetext{b}{Angular diameter estimate from deconvolved major axis of dust emission Gaussian fit (Table \ref{sizes}) using the 
Band 7 measurements for Arp 220 (since the Band 9 dust emission is optically thick) and Band 9 for NGC 6240 (since the SNR is better in Band 9).}
\tablenotetext{c}{Mean gas surface density and H$_2$ column density assuming the gas mass is distributed uniformly over a disk with the radius given in column 8. }
\tablenotetext{d}{Rayleigh-Jeans correction factor from Equation \ref{flux}. }
\end{deluxetable*}\label{masses}

\begin{figure*}[t]
\plotone{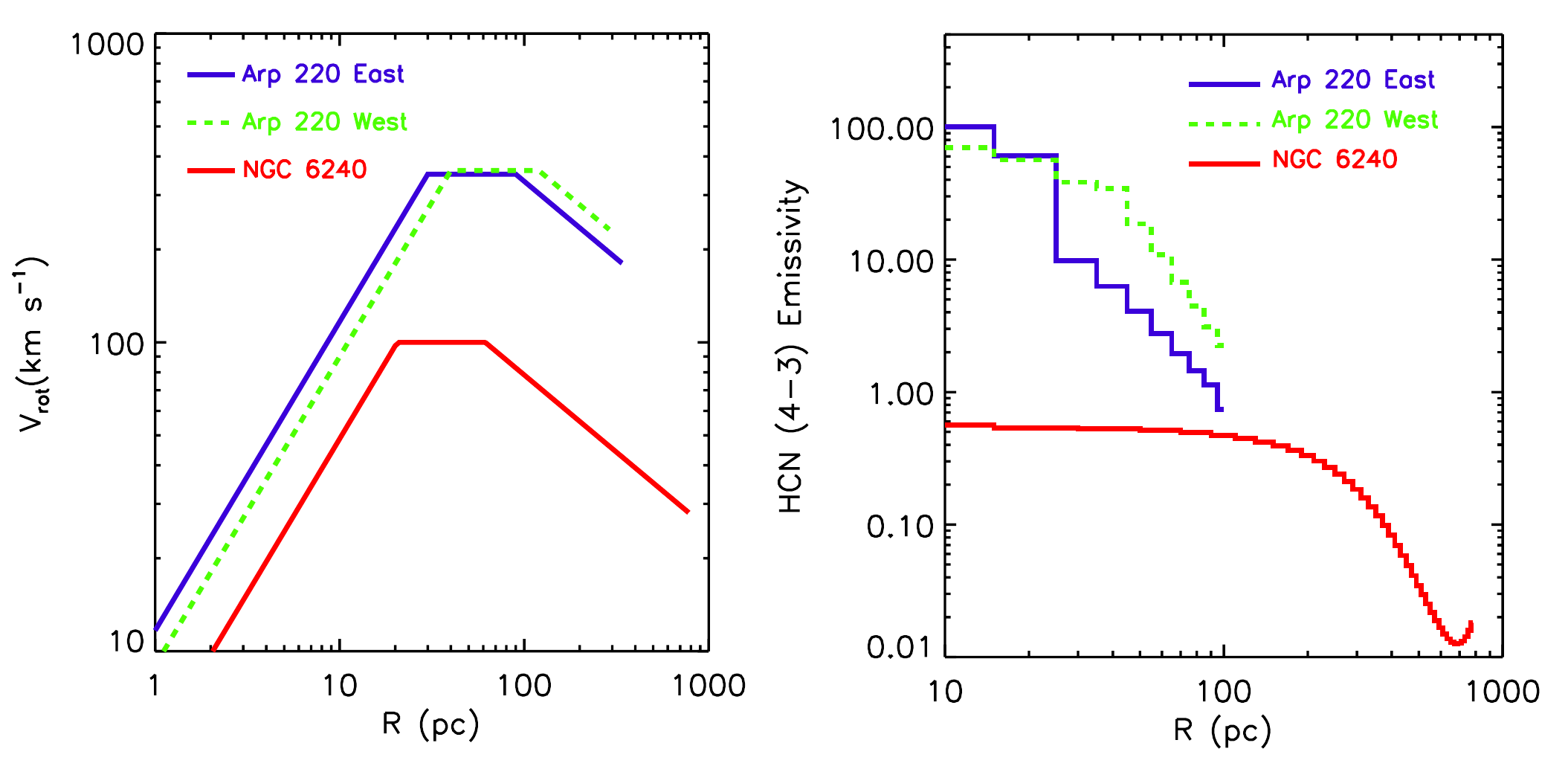}
\caption{The rotation velocity (left) and emissivity radial distributions (right) derived from the maximum likelihood kinematic deconvolution for the nuclear disks in Arp 220 East and West 
and NGC 6240.} 
\label{model_emiss} 
\end{figure*}

\begin{figure*}[t]
\plotone{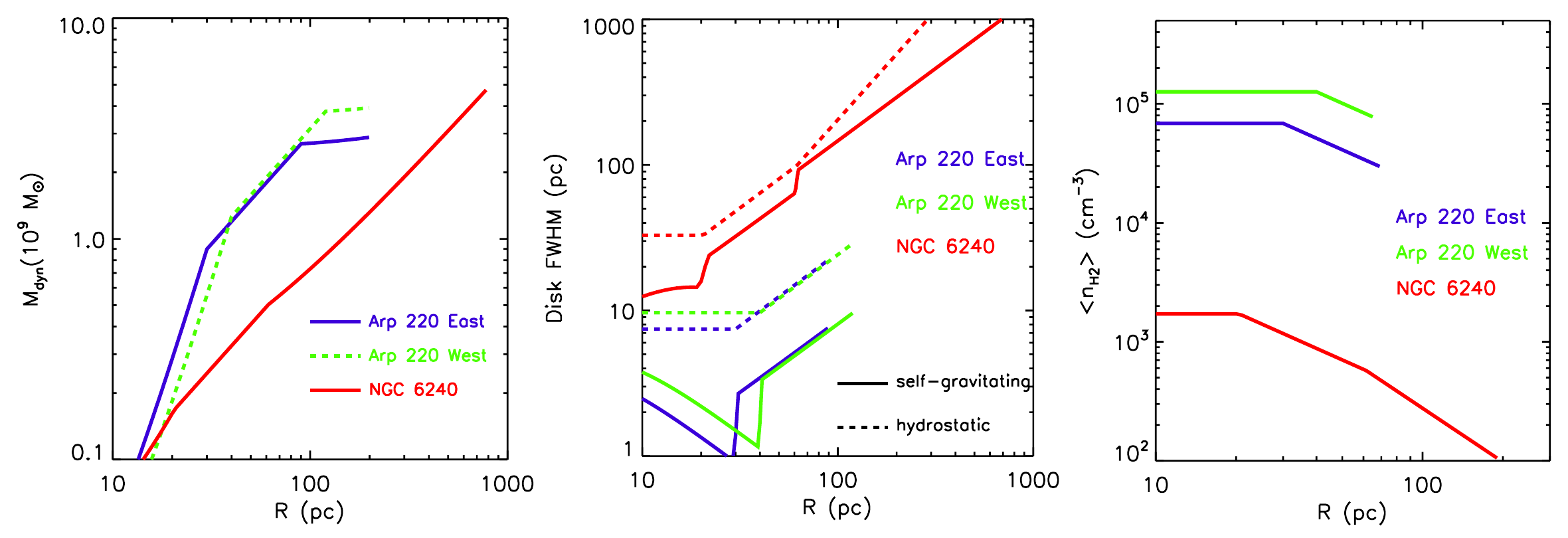}
\caption{The dynamical mass (Equation \ref{mdyne})(left), disk thickness (middle -- Equations \ref{dz1}  and \ref{dz2}) and mean disk gas densities (right) are shown, based on the rotation velocity and velocity dispersions derived from the maximum likelihood kinematic deconvolution for the nuclear disks in Arp 220 East and West 
and NGC 6240. The mean disk gas densities are shown for the ISM masses from Table \ref{masses}, distributed with constant mass surface density out to the radius given in 
Table \ref{masses}. The scale height as a function of radius is that of the hydrostatic case (solid lines in Figure \ref{physical}, obtained using Equation \ref{dz1}). The mass surface density 
was divided by a factor 1.36 to remove the He mass contribution.} 
\label{physical} 
\end{figure*}

\section{Nuclear Disk Models}\label{nuclear_disks}

\begin{deluxetable}{lrrrrr}
\tabletypesize{}
\tablecaption{\bf{Disk Models}} 
\tablewidth{0pt}
\tablehead{\colhead{Source}  & \colhead{$V_0$} & \colhead{$R_0$}   & \colhead{incl.} & \colhead{PA}   & \colhead{$\sigma_{v}$} \\
\colhead{}    & \colhead{} &  \colhead{}}
\startdata
&  \kms & pc & \deg & \deg & \kms \\
\hline \\
Arp 220 East &      350   &    30 &   71  & 47 & 90 \\
Arp 220 West &       360   &   40 &   64 & -15 & 90 \\
NGC 6240 &  100    &   20 &  70 & -6 &  160 \\
\enddata
\end{deluxetable}\label{models_table}

In Arp 220, the molecular gas and dust is likely in disk-like structures, centered on each nucleus of the 
merging galaxies, separated by only 1\arcsec or a projected separation of 361 pc. Although the direct imaging does not 
conclusively show the large ratio of major/minor axes expected for an edge-on disk, the monotonic velocity gradients along the
major axes of both structures (see Figure \ref{arp_line_sv}) are suggestive of rotating disks \citep[as seen here and earlier by][]{sak99,sak09}. Disk-like 
configurations for the nuclear gas concentrations in merging systems is also expected theoretically \citep{bar96}, 
given the highly dissipative nature of the dense molecular gas damps motion along the rotation axis. Our conclusion that  the gas constitutes a significant fraction of the 
total mass in the nuclei is also to be expected in the late stages of merging since the gas, being more dissipative than the
pre-existing stellar systems, will sink to smaller radii more quickly and hence lead to a large gas mass fraction in the nuclei. In NGC 6240, which is probably somewhat less advanced in the
merging process, the gas structure peaks near but not on the southern nucleus with a tidal bridge extending toward the northern nucleus. In NGC 6240,  the observed kinematics are not
ordered with a gradient along a single major axis.

In Appendix \ref{app}, we describe a kinematic deconvolution technique originally developed by \cite{sco83} which yields a maximum likelihood 
fit of a simple disk model to the observed HCN line profiles mapped over the area of the disk. The fit obtains a simultaneous solution for a parametric rotation curve 
and an axisymmetric, radial emissitivity distribution which best fits the observed line profiles seen in HCN (4 - 3) 
across Arp 220 East and West and NGC 6240. The objective of this modeling exercise is to yield a self-consistent rotation curve and emissivity distribution which agrees with the observed line profiles, and to see if this dynamic model yields a total dynamical mass consistent with the ISM masses estimated from the dust emission.

The  parameters to be solved for are: the velocity dispersion $\sigma_v$ 
of the gas; the inclination i of the disk; the PA of the major axis of the disk and the radial emissivity function. The radial emissivity function was taken to be a step function in radius (R)
with 30 equal width bins in R with no imposed continuity between adjacent radial bins. Positions and velocities were measured relative to the spatial and velocity 
centroids of the HCN emission; specifically, the centroid position was taken to be R=0 for the nucleus and the centroid velocity was adopted as the systemic velocity of each disk. These spatial offsets 
were in all cases $\leq 0.05$\arcsec from the dust continuum peaks and the velocity offsets were $\leq 20$ \kms from the systemic velocities.

The minimized $\chi^2$ parameters are given in Table \ref{models_table} and the rotation curves and radial emissivity distributions are shown in Figure \ref{model_emiss}. 
 In both Arp 220 East and West the general agreement between the model and observed spectra
is acceptable given the simplified assumptions of the model -- axisymmetry and no radiative transfer considerations. In both Arp 220 nuclei, the structures are consistent with being rotation dominated. 

In 
NGC 6240 the line profiles fits are also good but here the dynamics are dispersion-dominated and the disk structure is therefore not as well constrained. 
\cite{med11} constrained the black hole mass is the southern nucleus of NGC 6240 using NIR AO measurements of the stellar kinematics 
using the CO bandheads and 2.3$\mu$m. Their range of acceptable masses is 0.9 - 2 $\times10^9$ \msun.  This mass is significantly above that indicated by the 
rotation curve shown in Figure \ref{physical}-left. The analytic rotation curve used in our fitting did not include a term for a point mass but even if it had included 
a Keplerian falloff at small radii, the HCN emission is not very centrally concentrated; the modeling cannot therefore constrain the rotation curve there.

 The derived emissivity distributions in both nuclei of Arp 220 
are strongly concentrated to very small radii, $R \lesssim 20$ pc;  in NGC 6240 the emission apparently arises fairly uniformly out to radii $\sim 250$ pc. For Arp 220 the 
instrumental resolution (HPBW) of the HCN (4 - 3) observation is $\sim 0.5$\arcsec ~ or a diameter of 180 pc. Thus the kinematic deconvolution is revealing a source $\sim 5$ times smaller 
and it will be interesting to see the results of higher resolution ALMA imaging in future cycles. 

The kinematics derived from the modeling can be used to estimate dynamical masses for these compact gas concentrations. 
For the combination of circular orbital motions and uniform random motions, an approximate dynamical mass is obtained from 
 \begin{eqnarray}\label{mdyne}
 M(R) &=& 2.3\times10^8 \left(V_{rot~100}^2 + \sigma_{v~100}^2\right)  R_{100} ~\msun  .
 \end{eqnarray}
 
\noindent The velocities are normalized to 100 \kms ~and the radius to 100 pc. For both nuclei in Arp 220 the dynamical masses are $\sim7\times10^8$ \msun ~at R = 20 pc and $\sim3 \times10^9$ \msun ~at R = 100 pc. 
For NGC 6240, the dynamical mass estimates are $\sim6\times10^8$ and $1.4\times10^9$ \msun ~at R = 250 and 600 pc.  
The vertical structure of the disks is a Gaussian function with thickness (FWHM) given by 
 \begin{eqnarray}\label{dz1}
\Delta z_{FWHM}(R) &\simeq& 100 \sigma_{v~100}  \left( {R_{100} \over{ V_{rot~100} }} \right) \rm pc. 
 \end{eqnarray}
 
 A more precise formulation of the disk thickness is not warranted at this time since it would require knowledge of the relative fractions 
 of the mass in stars and gas and a more accurate rotation curve. In the event that the disk is entirely self-gravitating 
 and the gas mass is completely dominant then it can be shown that the thickness depends quadratically on the velocity dispersion and is given by 
  \begin{eqnarray}\label{dz2}
  \Delta z_{FWHM}(R) &\simeq& {0.233 \sigma_v^2 \over{G \Sigma}}  ~  \nonumber  \\
 &=& {53 \sigma_{v~100}^2 \over{ \Sigma_{10^4}}}  \rm pc 
 \end{eqnarray} 
 
\noindent where $\Sigma$ is the local mass surface density of the disk, here normalized to $10^4 \msun \rm pc^{-2}$ \citep[see][]{spi42,sco97}. 
 
 Figure \ref{physical} shows the dynamical mass, disk thicknesses (from both Equations \ref{dz1} and \ref{dz2}) and mean disk density as functions of radius 
 using the rotation curves from Figure \ref{model_emiss}. The density is calculated from the dynamical mass as a function of radius  and expressed in terms of equivalent volume density of H$_2$ 
 (in anticipation of the possibility that the gas makes up a large fraction of the overall mass). 

\bigskip

\section{High Excitation Gas traced in HCN and CS}\label{dense_gas}

In the Arp 220 nuclei, both the HCN (4 - 3) and CS (7 - 6) emission lines are optically thick but subthermally excited.\footnote{We are not aware of $^{13}$C isotope
detections for HCN or CS in Arp 220 but in virtually all Galactic sources with significant HCN and CS emission, the lines are optically thick since the 
rare isotope molecular emissions are much stronger, relative to the abundant isotope, than the isotopic abundance ratios.} Their peak brightness temperatures 
(Table \ref{fits}) are 20 - 40 K (HCN) and 7 - 10 K (CS), implying that they are excited well above the 2.7 K CMB, yet they are significantly less 
than the CO brightness temperatures and the $\sim100$ K dust temperatures for the resolutions probed here. In this subsection 
we provide an analytic treatment of the molecular excitation to constrain the physical conditions in the line emission regions. 

In the sub-thermal, optically thick regime of excitation, the molecular emission is provided by two processes: collisional excitation by collisions 
with H$_2$ and line photon radiative trapping \citep{sco74,gol74}. Here, we extend the 2-level analytic treatment in \cite{sco74} for our analysis of the HCN (4 - 3) emission in Arp 220. Although one could perform a full multi-level statistical equilibrium calculation for the level 
populations, this analytic treatment provides better physical insight and yields the dependences on the key parameters: 
$n_{H_2}$ (the H$_2$ volume density), $X_{m}=n_{m}/n_{H_2}$ (the molecular abundance relative to H$_2$) and T$_k$ (the gas temperature). (A similar analytic approach is given by \cite{gol12} for the [CII] fine structure line.)

\subsection{Two Level Excitation with Photon Trapping}

We consider a 2-level molecular system (e.g. HCN J = 3 and 4) in which the level populations are determined by a balance of collisions with H$_2$, spontaneous decay and line 
photon absorption and stimulated emission with $\tau > 1$. The ratio of upper to lower level populations is then given by 
 \begin{eqnarray} \label{e1}
 {n_u \over{n_l}}  &=& {n_{H_2} \left<\sigma v\right>_{ul} g_{ul} e^{-h\nu/kT_k} \over{n_{H_2} \left<\sigma v\right>_{ul} + A_{ul}/\tau }} . 
 \end{eqnarray}
 
 \noindent $g_{ul}$ is the ratio of statistical weights ($=g_u/g_l$), $\left<\sigma v\right>_{ul}$ is the downward collision rate coefficient and $T_k$ is the gas kinetic temperature. In Equation \ref{e1}, the spontaneous decay rate $A_{ul}$ has been reduced by the photon escape probability, $\beta = (1-e^{-\tau})/\tau  \simeq 1/ \tau $ for $\tau > 1$. 
The excitation temperature ($T_{ul}$) characterizing the level populations is then 

 \begin{eqnarray}
{ T_{ul} \over {T_k }}  &=& {1 \over{ 1 + {T_k \over{T_0}} ~ ln \left(1 + \chi \right) }} ~   \label{e2} \\
\rm where ~ \chi &=& {A_{ul}/\tau \over{ n_{H_2} \left<\sigma v\right>_{ul}}}   \label{e3}
 \end{eqnarray}
  
\noindent and $T_0 = h\nu / k$. The line optical depth is given by
 \begin{eqnarray}
 \tau  &=& A_{ul} { g_{ul} c^3 \over{8\pi  \nu^3 ~ dv/dr }} n_l \left(1-e^{-T_0/T_{ul}}\right)  ~  . 
 \label{e4}
 \end{eqnarray}
  \noindent Here, $dv/dr$ is the line-of-sight velocity gradient.

 Combining Equations \ref{e3} and \ref{e4} one sees that the excitation temperature in an optically thick transition is in fact \emph{independent of the spontaneous decay 
 rate} (A$_{ul}$). This is because transitions with higher spontaneous decay rates also have proportionally higher optical depth and hence lower photon escape probabilities. Although this has been shown before \citep{sco74,gol74}, we restate this result here since it is often asserted that the greater dipole moment molecules have a higher critical density due to their more rapid spontaneous decay -- this is not physically correct for optically thick transitions such as those of HCN and CS. In fact, the excitation temperature is independent of the A-coefficient; hence, the line brightness 
 temperature depends only on the factor $n_{H_2} n_{m} / (dv/dr)$ where $n_m$ is the volume density of the molecules (reflected above in $n_l$).

 To proceed further, we make the approximation that the rotational level populations can be characterized by a single excitation temperature $T_x$. This excitation temperature can then be used to relate the 
lower level population n$_l$) to the total molecular density ($n_m$) via a Boltzmann distribution. For a linear molecule, the rotational partition function is $Z = k T_x / (hB)$ where B is the rotation constant. Thus,  
 \begin{eqnarray}
n_{l}   &=& g_l e^{-E_l/kT_x} n_m / Z  ~  \nonumber \\
&=& {g_l e^{-E_l/kT_x} n_m hB \over{ k T_x}} . \label{e5}
 \end{eqnarray}
 
For linear molecules like HCN and CS, the transition J $\rightarrow$ J - 1 has $E_l = hB (J^2-J)$ and $T_0 = h\nu/k = 2JhB/k$ (B = 44.32 and 24.50 GHz for HCN and CS). Equation \ref{e2} 
 then becomes
  \begin{eqnarray}
{ T_{x} \over {T_k }}  &=& {1 \over{ 1 + {T_k \over{T_0}} ~ ln \left(1 + \psi ~f(T_x) \right) }} ~   \label{e6} \\
\rm where ~ \psi &=& { 64\pi  (JB/c)^3  \over{ (n_m / (dv/dr)) ~ n_{H_2} ~ g_u \left<\sigma v\right>_{ul}}}  \label{e7} \\
 \rm and ~  f(T_x) &=& {k T_x / hB \over{e^{-hB(J^2-J)/kT_x} - e^{-hB(J^2+J)/kT_x} }} .~ \label{e8} 
  \end{eqnarray}
  
  \begin{figure*}[t]
\plottwo{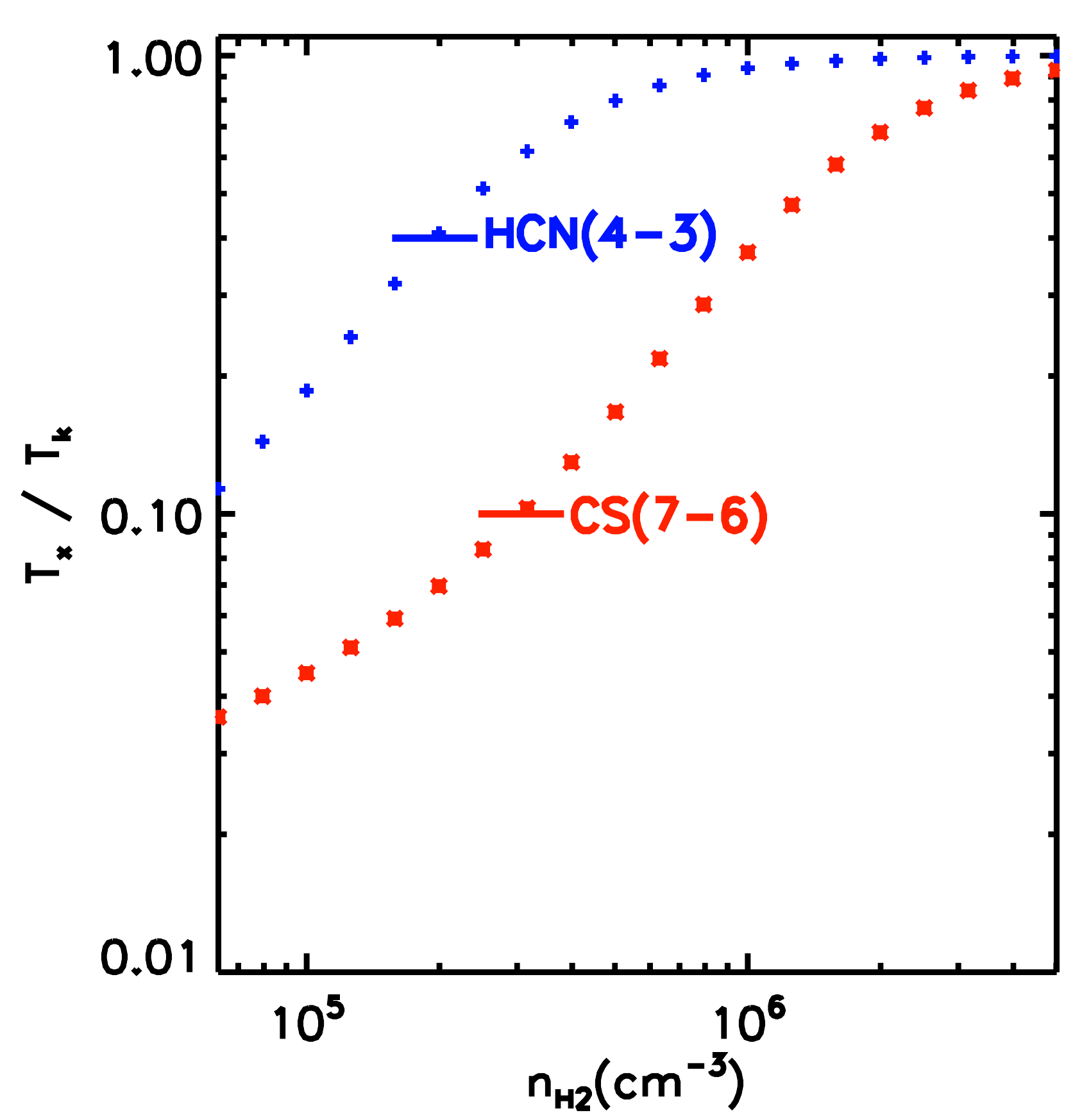}{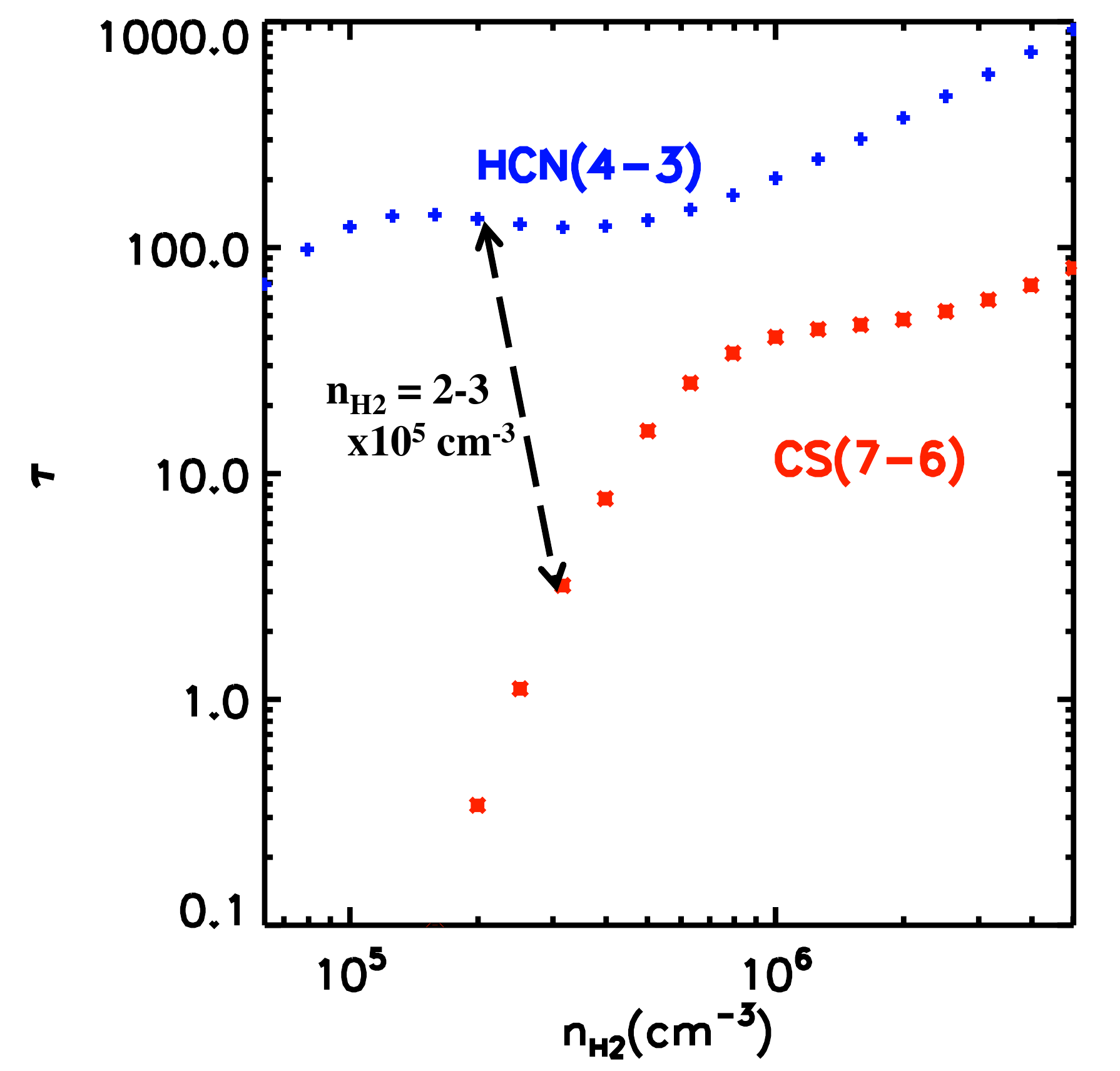}
\caption{Left -- The ratio of the level excitation temperature to the gas kinetic temperature ($T_x/T_k$) is shown as a function of density for the HCN (4 -3) and CS (7 - 6) lines, calculated from Equation \ref{e6} for the parameters discussed 
in the text, specifically $T_k = 100$ K and standard galactic abundances $2\times10^{-8}$ and $2\times10^{-9}$ for HCN and CS relative to H$_2$.  
The observed thermalization ratios (derived from the line brightness temperatures, see text) are $\sim0.4$ and 0.1, requiring $n_{H_2} = $2.0 and 3.0$\times10^5$ cm$^{-3}$, respectively for HCN and CS. Right -- Line optical depths are shown. The optical depths corresponding to the solution densities are $\sim130$ and 3 for the HCN (4 -3) and CS (7 - 6) lines, averaged over the line profile.} 
\label{hcn_cs_excite} 
\end{figure*}

  \subsection{HCN (4 - 3) Excitation in Arp 220}
  
This 2-level formalism can be applied to the observed HCN (4 - 3) emission in Arp 220 to constrain the gas density in the nuclear sources. In the nuclei at 
radii less than $\sim65$ pc we adopt a gas kinetic temperature T$_k = 100$ K, based on the dust blackbody-limit temperature and the expectation that at densities 
above $10^4$ cm$^{-3}$ the gas should be thermally coupled to the dust. This high kinetic temperature is consistent with the brightness temperatures 
observed for CO \citep{sak09}. For HCN (4 - 3), the peak line brightness temperatures are $\sim20$ and 40 K (Table \ref{sizes}). Assuming the line is optically 
thick (as in Galactic sources), these brightness temperatures imply excitation temperatures $T_x \sim 20$ and 40 K. In the evaluation below, we adopt 
 $T_x \simeq$ 40 K and therefore the gas density will be constrained to that giving $T_{x} / {T_k } \simeq 0.4$ in Equation \ref{e6}. 
 
For HCN, $hB/k = 2.13$ K and for the J = 4 -3 transition, $T_0 = 17.0$ K with $A_{4-3} = 2.1\times10^{-3}$ sec$^{-1}$.  The collision rate coefficients for HCN J=4 were taken from the He-HCN rate coefficients 
calculated by \cite{dum10}, scaled by a factor 1.36 to go to HCN-H$_2$\citep[see][]{gre76}. For this analytic analysis, it is appropriate to use the sum 
of the collision rate coefficients out of the J=4 upper level to any other rotational level (both below and above J=4) since all of these 
transitions couple the level to the gas kinetic temperature. This summed rate coefficient is $\left<\sigma v\right>_{J=4} = 2.68\times10^{-10} (T_k / 100)^{-0.22}$ for $T_k \simeq 50 - 150$K (using the tables given in \cite{sch05a}). 
 
For the characteristic line-of-sight velocity gradient, we divide the line FHWM ($\sim500$ \kms) by the diameter of the nuclear sources (130 pc, Table \ref{masses}),  yielding $dv/dr \sim 4 $\kms pc$^{-1}$ ($=1.3\times10^{-13} \rm sec^{-1}$). Lastly, we set the abundance of HCN relative to H$_2$ to the typical value obtained 
for Galactic sources \citep[$X_{HCN} \sim 2\times10^{-8}$ -- ][]{ber96,lah00} so that the molecular volume density is $n_m = X_{HCN} n_{H_2}$.

  \subsection{CS (7 - 6) Excitation in Arp 220}
  
 For CS (7 - 6), the peak line brightness temperatures are $\sim7.5$ and 10 K (Table \ref{sizes}), i.e. about 1/4 of those for HCN (4 - 3), implying that $T_{x} / {T_k } \simeq 0.1$ in Equation \ref{e6}. 
 For CS, $hB/k = 1.175$ K and for the J = 7 - 6 transition, $T_0 = 16.5$ K with $A_{7-6} = 8.4\times10^{-4}$ sec$^{-1}$.  The collision rate coefficients for CS \citep{sch05a} have total $\left<\sigma v\right>_{J=7} = 2.87\times10^{-10} (T_k / 100)^{0.15}$, once again summing all rate coefficients 
 from  J=7. We use the same line-of-sight velocity gradient as for HCN, i.e. $dv/dr =1.3\times10^{-13} \rm sec^{-1}$. For CS, the Galactic and nearby galaxies abundance is typically $X_{CS} \sim 5\times10^{-9}$ \citep{pag95,shi03,ber97}.

  \subsection{Densities required for the Observed Emission}\label{density_ex}
  
Figure \ref{hcn_cs_excite}-Left shows the resulting ratios of $T_x / T_k$ obtained for HCN (4 - 3) and CS (7 - 6) from Equation \ref{e6} as a function of density. 
The observed fiducial values for $T_x / T_k$ are $\sim0.4$ and 0.1 in Arp 220 West for HCN and CS, respectively -- these are shown by the horizontal bars. For the adopted molecular abundances 
the required densities are $n_{H_2} = 2.0$ and $3.0 \times10^5$ cm$^{-3}$ for HCN and CS. The densities derived from this analytic 2-level 
treatment are within 10\% of those derived using a full multilevel code such as RADEX. Figure \ref{hcn_cs_excite}-Right shows the line optical depths as a function of density, indicating that both lines are optically thick ($\tau = 130$ and 3) --  the HCN line optical depth is over 100!

It is noteworthy that these two lines indicate fairly consistent gas volume densities (2.0 and $3.0 \times10^5$ cm$^{-3}$) without requiring abnormal molecular abundances -- so we conclude that  $n_{H_2} \sim 2\times10^5$ cm$^{-3}$. This volume density for the line emitting gas is only slightly higher than the mean volume density ($\sim 10^5$ cm$^{-3}$, estimated from the dynamical mass and from the dust-based mass estimate). The fact that these agree 
so closely implies the volume is almost entirely filled with gas at density $1-2\times10^{5}$ cm$^{-3}$. Thus, \emph{ the nuclear gas configuration must be 
quite uniform and not a cloudy or swiss cheese -like structure}. 

The above analysis considers only collisional excitation. In some instances, the excited rotational levels can be populated via absorption of 
near/mid infrared dust continuum photons in the molecular vibrational bands, followed by spontaneous decay to the excited rotational levels of the ground 
vibrational state \citep[][]{car81}. For CS and HCN, this mode starts for dust black body temperatures above 
$\sim$120 and 400 respectively for CS and HCN but considerably higher $T_d$ are required for the high J states observed here \citep[see][]{car81}.

\section{Discussion}

Here, we briefly discuss the constraints on variations in the CO conversion factor used to translate CO (1 - 0) emission line 
flux to mass of H$_2$ and the estimation of star formation rates from the IR fine structure lines such as CII.

\bigskip

\subsection{CO-to-H$_2$ conversion factor}\label{conversion}

\begin{deluxetable}{lccccr}
\tabletypesize{}
\tablecaption{\bf{Arp 220 -- CO and Dust Masses}}
\tablewidth{0pt}
\tablehead{\colhead{}  & \colhead{$S_{2-1} \Delta V$} & \colhead{$S_{1-0} \Delta V$}    & \colhead{$ISM(CO)$}   & \colhead{$ISM(dust)$} }
\startdata
&  & & $10^9$\msun & $10^9$\msun & \\
\hline \\
East &      120   &    30 --   46  & 3.9 -- 6.0 &  2.0 \\
West &       187   &   47 --   72 & 6.0 -- 9.2 &  4.2 \\
\enddata
\tablecomments{CO (2 - 1) fluxes for the nuclei are from \cite{sak99}. Similar 2-1 fluxes are given by \cite{dow98}: 130 and 220 Jy \kms (assuming their table entries for 
the East and West nuclei are reversed). Flux units are Jy \kms. A lower limit on the 1 - 0 flux is obtained by assuming the flux scales as $\nu^2$ and the ratio is therefore 4:1. The upper limit is obtained by adopting the 
global CO (2-1) / (1-0) flux ratio of Arp 220 \citep[1071/410 = 2.6][]{sco97}. ISM(CO) is the ISM mass estimate obtained from the CO (1-0) 
flux assuming the standard Galactic CO-to-H$2$ conversion factor and ISM(dust) is the ISM masses estimated from dust RJ continuum flux as listed in Table \ref{masses}.} \label{co_h2_tab}
\end{deluxetable}

Using the masses derived from the dust emission in \S \ref{dust_mass} and Table \ref{masses}, we can 
assess the effective CO (1 - 0) conversion factor for the Arp 220 nuclei (assuming the masses obtained 
from the dust emission are valid). To circumvent the problem that angular resolution of existing CO (1 -0) is insufficient to separate the two nuclei, and to separate their emission from the more extended CO emission, we make use of the 
CO (2-1) observations of \cite{sak99} at 0.5\arcsec~ resolution. 
To estimate the effective (1-0) flux one can scale the (2-1) flux by a factor 1/4, assuming the two lines are thermalized and have constant brightness temperature. Alternatively, 
one could scale by the global (2-1)/(1-0) flux ratio = 2.6 \citep{sco97}. The 1-0 flux obtained using this latter ratio should be viewed as an upper limit since the 
global flux ratio includes low excitation CO from larger galactic radii. These estimates are listed in Table \ref{co_h2_tab}.

For the standard Galactic CO-to-H$_2$ conversion factor $\alpha_{CO (1-0)} = 3\times10^{20}$ H$_2$ cm$^{-2}$ (K \kms)$^{-1}$, the H$_2$ mass is 
M$_{H_2} = 1.18\times10^4 S_{1-0} \Delta V d_{Mpc}^2$ \citep[see Appendix in][]{san91}. The resulting CO-based mass estimates are given 
in column (4) of Table \ref{co_h2_tab} after multiplying by a factor 1.36 to account for He. The ISM mass estimates from the CO (1 - 0) using the 
standard Galactic CO-to-H$_2$ conversion factor are 2 - 3 times higher than the estimates obtained from the dust continuum (see Table \ref{co_h2_tab}). 
The lower CO estimates are probably more valid for the nuclei since the flux ratio of the high excitation gas is very likely to be thermalized and therefore have the limiting  4:1 ratio.

For self-gravitating molecular clouds or galactic nuclei where the gas is the dominant mass component, one expects the CO-to-H$_2$ 
conversion factor to scale as $\sqrt{n_{H_2}} / T_k $ \citep{dic86,bry96}, relative to the density and temperature of the Galactic GMCs for which the 
standard conversion factor was derived. The Galactic GMC ratio is $\sim\sqrt{300} / 10 \sim 1.7$ and the ratio for the Arp 220 nuclei is 
$\sim\sqrt{10^5} / 100 \sim 3.2$. Thus a factor $\sim2$ change in $\alpha_{CO (1-0)}$ is to be expected -- this is consistent with the 
discrepancy between the CO- and dust-based ISM mass estimates given in Table \ref{co_h2_tab}. 

Thus we conclude 
that the effective CO-to-H$_2$ conversion factor in the Arp 220 nuclei is approximately a factor 2 reduced from the standard Galactic value. 
Most of this reduction can be attributed to the expected changes in the mean gas density and gas kinetic temperature. The 
CO-to-H$_2$ conversion factor can also be reduced if the line width is increased due to there being a significant stellar mass contribution in the self-gravitating 
region containing the molecular gas \citep{bry96,dow98}. In the Arp 220 nuclei, we find that this mechanism is not necessitated since the apparent changes in the conversion factor can be largely attributed to the expected scaling for higher T$_k$ and density. Lastly, we point out that the masses estimated here from the dust emission and from the CO (as discussed above) are consistent with the dynamical mass estimated from the kinematic modeling (as shown in Figure \ref{physical}-left). Thus the derived conversion factor, reduced by a factor 2 from the standard one, is certainly not implausible.

\subsection{SFRs from CII and the 'IR Line Deficits'}


The 158$\mu$m CII line is the most luminous far-infrared line from dusty star forming regions and a number of investigations have attempted to 
use the CII line fluxes to probe SFRs \citep[cf.][]{mal01,sta10}.  However, it is now well established that in many ULIRGs the CII line luminosity is suppressed 
by factors of 10-100 relative to the far infrared luminosity \citep{luh03,gra11,dia14}. The latter is a robust indicator of dust obscured SF in galaxies not having strong AGN.

There are a number of explanations which have been proposed to account for the suppressed line flux: 1) a sufficiently high density in the emission line regions 
that the fine structure levels are collisionally de-excited (i.e. the gas density is above the fine structure transition critical density), 2) some of the UV luminosity which would normally photoionize 
the HII and PDR gas is absorbed by dust, thus enhancing $L_{IR}$ relative to L$_{CII}$ or 3) emission line flux is suppressed by absorption due to overlying dust. 

For the CII line, the critical densities are $n_e \simeq50$ cm$^{-3}$ for ionized gas at 10$^4$ K and $n \simeq3000$ cm$^{-3}$ for neutral gas at 100 K. The mean gas densities 
estimated above for the Arp 220 nuclei ($n_{H_2} \simeq 5\times10^5$ cm$^{-3}$) are 100 times higher than the neutral gas critical density; thus  
in Arp 220, collisional suppression can fully account for the line deficit. In addition, the foreground dust opacities in Arp 220 are such that the dust is optically thick 
out to $\lambda \gtrapprox$ 300$\mu$m, so it is expected that there will be substantial dust absorption of the line emission, even if the intrinsic emission luminosity is normal. 


\cite{dia13} find that the CII line deficit is correlated with there being warm dust in the LIRGs and ULIRGs. They suggest that this is consistent with absorption 
by dust in the PDRs of the UV longward of the Lyman limit. This is the UV which is responsible for ionizing CII in the
PDR regions. 
Given the physical conditions in Arp 220, the CII line suppression is inevitable due 
to collisional de-excitaiton and the line emission will also be absorbed by overlying dust so it is not necessary to invoke this additional mechanism. It is important also to 
recognize that a significant fraction of the CII emission (25-50\%) probably arises from the HII regions (as opposed to PDRs) and this would not be affected by the PDR-dust. The HII region CII emission component would therefore be at its normal strength relative to the SFR; since this component is a significant fraction of the 
normal CII emission, it would then be difficult to reach line deficits reaching factors 10 - 100.

\subsection{Eddington Limit}

In nuclear SB and AGN sources with a dusty ISM, the absorption of radiation by the dust can produce an outward radiation 
pressure force comparable or larger than gravity, leading to self-regulation of the activity and radiatively driven outflows \citep{sco01,sco03,tho05,mur05}. Thus there exists an Eddington limit on the luminosity-to-mass ratio and above this limiting L/M
the ISM is driven outwards and hence the AGN and SB fueling is cut off. 

Given the rotational curve for a galactic nucleus, one can write the Eddington luminosity as 
  \begin{eqnarray}
L_{\rm Edd}&=&\frac{4\pi G M(<R) c}{\kappa_R} =\frac{4\pi V_{\rm rot}(R)^2\,Rc}{\kappa} \nonumber \\
&\simeq&1.5\times10^{12}\,{\rm L_\odot}\,\left(\frac{V_{\rm rot}}{350\,{\rm km\,s^{-1}}}\right)^2\left(\frac{R}{40\,{\rm pc}}\right) \nonumber \\
& & \times \left(\frac{10\,{\rm cm^2\,g^{-1}}}{\kappa_R}\right) \rm ~~ for ~Arp 220 \\
 \nonumber \\
\noindent \rm and  \nonumber \\
&\simeq& 3\times10^{11}\,{\rm L_\odot}\,\left(\frac{V_{\rm rot}}{100\,{\rm km\,s^{-1}}}\right)^2\left(\frac{R}{100\,{\rm pc}}\right) \nonumber \\
&& \times \left(\frac{10\,{\rm cm^2\,g^{-1}}}{\kappa_R}\right) \rm ~~ for ~NGC ~6240
\label{ledd}
  \end{eqnarray}
where we assumed a relatively high value of the Rosseland opacity $\kappa_R$ for the dust. 

The effective opacity for the dust to absorb the outflowing radiative momentum of course 
depends on the typical wavelength of the photons at each radius. The original luminosity from the AGN or SB is
in the optical-UV, but if the overlying dust is optically thick, these photons are absorbed in the first column with A$_V \sim 1$ mag 
(or N$_H = 2\times10^{21}$ cm$^{-2}$ for standard gas-to-dust abundance). The absorbed radiation is then re-emitted in the infrared -- at successively longer 
wavelengths at larger radii since the dust temperatures decreases at larger radii. 
The coupling of the dust to the radiation field depends on the effective dust opacity at the characteristic wavelength 
of the photons at each radius. The Eddington ratio, L/M or equivalently $L/(V^2/R)$, will therefore decrease at larger radii
in an optically thick dust cloud since the effective opacity is reduced for longer wavelength photons. \cite{sem03} give $\kappa_R(T)\simeq2\,{\rm cm^2\,g^{-1}}(T/100\,{\rm K})^2$ for $T\lesssim200$\,K and $\kappa_R(T)\simeq5\,{\rm cm^2\,g^{-1}}$ for $T\gtrsim200$\,K for a dust-to-gas ratio normalized to the Milky Way value.  

Prior treatments of the Eddington limit in the context of starbursts have lacked strong constraints on the radial profile of the radiation field to compare with the self-gravitating mass. 
The star formation rate $\simeq120$ M$_\odot$ yr$^{-1}$ for the West nucleus of Arp 220 based on the infrared luminosity.  We can convert this to a bolometric power by assuming a standard Kroupa IMF:
  \begin{eqnarray}
L_{\rm bol}&=&\epsilon ~SFR~ c^2   \\
&\simeq&1.1\times10^{12}
\,{\rm L_\odot}\,\left(\frac{SFR}{100\,{\rm M_\odot\,\,yr^{-1}}}\right)   \times \left(\frac{7\times10^{-4}}{\epsilon}\right) \nonumber
\label{ledd2}
  \end{eqnarray}
For Arp 220, the Eddington ratio is then near-unity for the Western nucleus and sub-Eddington for the Eastern nucleus by a factor of a few for the nominal $\kappa_R$.  

Alternatively, one could take the HCN (4-3) emissivity profiles and {\it assume} that $L_{\rm bol}(R)$ tracks the HCN emissivity: $L_{\rm bol}\propto L_{\rm HCN(4-3)}$ (Figure \ref{model_emiss}) and then take the dynamical mass as a function of radius from Figure \ref{physical}. The emissivity is rising rapidly at smaller radii in the nuclei of Arp 220, but the enclosed mass must be falling.  Thus, the Eddington ratio should be rising toward the core of the nuclei -- ultimately leading to an inner radius where the Eddington ratio 
exceeds unity and material would be ejected. Given the uncertainties in the mass and emissivity distributions derived here; we defer 
of numerical analysis of this approach to a later dataset with higher spatial resolution. 
 
\section{Summary and Perspective}

These ALMA data probing the high excitation gas in the nuclei of two luminous IR galaxies clearly reveal the spectacular gas concentrations and physical 
conditions in these late stage mergers. Consistent values are obtained for the gas mass derived from the Rayleigh-Jeans dust emission and 
for the dynamical mass derived from modeling the observed emission line profiles. 

The nuclear disks in Arp 220 are found to be extraordinarily compact with 
radii $\leq 65$ pc and masses $\sim2\times10^9$ \msun. The typical disk thickness $\sim10$ pc is estimated from the line profile velocity dispersion relative to the 
rotation velocities. The gas densities derived from the excitation requirements of the HCN and CS emission are consistent, yielding $n_{H_2} = $2$\times10^5$ cm$^{-3}$. Since these densities are within a factor two of the volume-average density for the nuclear gas, this high density gas must uniformly fill 
the volume and not be cloudy or swiss cheese -like (see \S \ref{density_ex}). 

There has been long standing uncertainty
with respect to using the CO emission from ULIRGs to estimate gas masses since the gas may be hotter and denser -- these changes affect the CO-to-H2
conversion factor in opposite directions and therefore reduce the shifts in $\alpha_{CO}$. The mm-lines can also have enhanced emissivity per unit gas mass if the central potential well of the 
galaxy nucleus is broadening the line emission. In the future, the effects can be resolved with high resolution ALMA imaging -- measuring the CO lines 
and dust emission as a function of radii to differentiate the different dependences. 

\bigskip 
Most of the analysis in this paper has focused on Arp 220 -- it is brighter and hence the data has higher signal-to-noise ratio and its structure, consisting of 
two disks centered on the near infrared nuclei, is simpler. On the other hand, it is fair to ask: why is NGC 6240 less luminous than Arp 220 while its twin AGNs are more active? Is this simply due to the incredibly high column densities ($> 10^{24-25}$ cm$^{-2}$) in Arp 220 which would preclude 
easy detection of an X-ray AGN there \citep[see][]{pag14}, Or is it the difference in total gas content, orbital configuration and/or mass distribution, gas affecting the AGN activity? The observations and modeling presented 
here clearly show a 3-4 times lower gas content in NGC 6240 with 3 times larger radius (see Table \ref{masses}). In addition, most of this ISM is
associated with the southern nucleus in NGC 6240 as compared with a more equal distribution in Arp 220.  Despite the above differences, 
both systems, with nuclei separated by $\sim$361 to 713 pc (projected),
are clearly on a precipitous 'death' spiral to coalesence in next $\sim$10-20 Myrs and subsequent rebirth as a more massive single galaxy. 

\acknowledgments
We would like to thank the referee for a very thorough reading of the manuscript. We thank Zara Scoville for proofreading the manuscript. 
KS is supported by the National Radio Astronomy Observatory, which is a facility of the National Science Foundation operated under cooperative 
agreement by Associated Universities, Inc. This paper makes use of the following ALMA data: ADS/JAO.ALMA\# 2011.0.00175.S. ALMA is a partnership of ESO (representing its member states), NSF (USA) and NINS (Japan), together with NRC (Canada) and NSC and ASIAA (Taiwan), in cooperation with the Republic of Chile. The Joint ALMA Observatory is operated by ESO, AUI/NRAO and NAOJ


\bibliography{scoville_ulirgs}{}

\begin{thebibliography}{}
\expandafter\ifx\csname natexlab\endcsname\relax\def\natexlab#1{#1}\fi

\bibitem[{{Aalto} {et~al.}(2009){Aalto}, {Wilner}, {Spaans}, {Wiedner},
  {Sakamoto}, {Black}, \& {Caldas}}]{aal09}
{Aalto}, S., {Wilner}, D., {Spaans}, M., {et~al.} 2009, \aap, 493, 481

\bibitem[{{Ant{\'o}n} {et~al.}(2004){Ant{\'o}n}, {Browne}, {March{\~a}},
  {Bondi}, \& {Polatidis}}]{ant04}
{Ant{\'o}n}, S., {Browne}, I.~W.~A., {March{\~a}}, M.~J.~M., {Bondi}, M., \&
  {Polatidis}, A. 2004, \mnras, 352, 673

\bibitem[{{Barcos-Mu{\~n}oz} {et~al.}(2014){Barcos-Mu{\~n}oz}, {Leroy},
  {Evans}, {Privon}, {Armus}, {Condon}, {Mazzarella}, {Meier}, {Momjian},
  {Murphy}, {Ott}, {Reichardt}, {Sakamoto}, {Sanders}, {Schinnerer},
  {Stierwalt}, {Surace}, {Thompson}, \& {Walter}}]{bar14}
{Barcos-Mu{\~n}oz}, L., {Leroy}, A.~K., {Evans}, A.~S., {et~al.} 2014, ArXiv
  e-prints, arXiv:1411.0932

\bibitem[{{Barnes} \& {Hernquist}(1992)}]{bar92}
{Barnes}, J.~E., \& {Hernquist}, L. 1992, \araa, 30, 705

\bibitem[{{Barnes} \& {Hernquist}(1996)}]{bar96}
---. 1996, \apj, 471, 115

\bibitem[{{Bergin} {et~al.}(1997){Bergin}, {Goldsmith}, {Snell}, \&
  {Langer}}]{ber97}
{Bergin}, E.~A., {Goldsmith}, P.~F., {Snell}, R.~L., \& {Langer}, W.~D. 1997,
  \apj, 482, 285

\bibitem[{{Bergin} {et~al.}(1996){Bergin}, {Snell}, \& {Goldsmith}}]{ber96}
{Bergin}, E.~A., {Snell}, R.~L., \& {Goldsmith}, P.~F. 1996, \apj, 460, 343

\bibitem[{{Bryant} \& {Scoville}(1996)}]{bry96}
{Bryant}, P.~M., \& {Scoville}, N.~Z. 1996, \apj, 457, 678

\bibitem[{{Bryant} \& {Scoville}(1999)}]{bry99}
---. 1999, \aj, 117, 2632

\bibitem[{{Carroll} \& {Goldsmith}(1981)}]{car81}
{Carroll}, T.~J., \& {Goldsmith}, P.~F. 1981, \apj, 245, 891

\bibitem[{{D{\'{\i}}az-Santos} {et~al.}(2013){D{\'{\i}}az-Santos}, {Armus},
  {Charmandaris}, {Stierwalt}, {Murphy}, {Haan}, {Inami}, {Malhotra},
  {Meijerink}, {Stacey}, {Petric}, {Evans}, {Veilleux}, {van der Werf}, {Lord},
  {Lu}, {Howell}, {Appleton}, {Mazzarella}, {Surace}, {Xu}, {Schulz},
  {Sanders}, {Bridge}, {Chan}, {Frayer}, {Iwasawa}, {Melbourne}, \&
  {Sturm}}]{dia13}
{D{\'{\i}}az-Santos}, T., {Armus}, L., {Charmandaris}, V., {et~al.} 2013, \apj,
  774, 68

\bibitem[{{D{\'{\i}}az-Santos} {et~al.}(2014){D{\'{\i}}az-Santos}, {Armus},
  {Charmandaris}, {Stacey}, {Murphy}, {Haan}, {Stierwalt}, {Malhotra},
  {Appleton}, {Inami}, {Magdis}, {Elbaz}, {Evans}, {Mazzarella}, {Surace}, {van
  der Werf}, {Xu}, {Lu}, {Meijerink}, {Howell}, {Petric}, {Veilleux}, \&
  {Sanders}}]{dia14}
---. 2014, \apjl, 788, L17

\bibitem[{{Dickman} {et~al.}(1986){Dickman}, {Snell}, \& {Schloerb}}]{dic86}
{Dickman}, R.~L., {Snell}, R.~L., \& {Schloerb}, F.~P. 1986, \apj, 309, 326

\bibitem[{{Downes} \& {Eckart}(2007)}]{dow07}
{Downes}, D., \& {Eckart}, A. 2007, \aap, 468, L57

\bibitem[{{Downes} \& {Solomon}(1998)}]{dow98}
{Downes}, D., \& {Solomon}, P.~M. 1998, \apj, 507, 615

\bibitem[{{Draine} {et~al.}(2007){Draine}, {Dale}, {Bendo}, {Gordon}, {Smith},
  {Armus}, {Engelbracht}, {Helou}, {Kennicutt}, {Li}, {Roussel}, {Walter},
  {Calzetti}, {Moustakas}, {Murphy}, {Rieke}, {Bot}, {Hollenbach}, {Sheth}, \&
  {Teplitz}}]{dra07b}
{Draine}, B.~T., {Dale}, D.~A., {Bendo}, G., {et~al.} 2007, \apj, 663, 866

\bibitem[{{Dumouchel} {et~al.}(2010){Dumouchel}, {Faure}, \& {Lique}}]{dum10}
{Dumouchel}, F., {Faure}, A., \& {Lique}, F. 2010, \mnras, 406, 2488

\bibitem[{{Dunne} {et~al.}(2000){Dunne}, {Eales}, {Edmunds}, {Ivison},
  {Alexander}, \& {Clements}}]{dun00}
{Dunne}, L., {Eales}, S., {Edmunds}, M., {et~al.} 2000, \mnras, 315, 115

\bibitem[{{Dunne} \& {Eales}(2001)}]{dun01}
{Dunne}, L., \& {Eales}, S.~A. 2001, \mnras, 327, 697

\bibitem[{{Engel} {et~al.}(2011){Engel}, {Davies}, {Genzel}, {Tacconi},
  {Sturm}, \& {Downes}}]{eng11}
{Engel}, H., {Davies}, R.~I., {Genzel}, R., {et~al.} 2011, \apj, 729, 58

\bibitem[{{Engel} {et~al.}(2010){Engel}, {Davies}, {Genzel}, {Tacconi},
  {Hicks}, {Sturm}, {Naab}, {Johansson}, {Karl}, {Max}, {Medling}, \& {van der
  Werf}}]{eng10}
---. 2010, \aap, 524, A56+

\bibitem[{{Goldreich} \& {Kwan}(1974)}]{gol74}
{Goldreich}, P., \& {Kwan}, J. 1974, \apj, 189, 441

\bibitem[{{Goldsmith} {et~al.}(2012){Goldsmith}, {Langer}, {Pineda}, \&
  {Velusamy}}]{gol12}
{Goldsmith}, P.~F., {Langer}, W.~D., {Pineda}, J.~L., \& {Velusamy}, T. 2012,
  \apjs, 203, 13

\bibitem[{{Graci{\'a}-Carpio} {et~al.}(2011){Graci{\'a}-Carpio}, {Sturm},
  {Hailey-Dunsheath}, {Fischer}, {Contursi}, {Poglitsch}, {Genzel},
  {Gonz{\'a}lez-Alfonso}, {Sternberg}, {Verma}, {Christopher}, {Davies},
  {Feuchtgruber}, {de Jong}, {Lutz}, \& {Tacconi}}]{gra11}
{Graci{\'a}-Carpio}, J., {Sturm}, E., {Hailey-Dunsheath}, S., {et~al.} 2011,
  \apjl, 728, L7+

\bibitem[{{Green} \& {Thaddeus}(1976)}]{gre76}
{Green}, S., \& {Thaddeus}, P. 1976, \apj, 205, 766

\bibitem[{{Greve} {et~al.}(2009){Greve}, {Papadopoulos}, {Gao}, \&
  {Radford}}]{gre09}
{Greve}, T.~R., {Papadopoulos}, P.~P., {Gao}, Y., \& {Radford}, S.~J.~E. 2009,
  \apj, 692, 1432

\bibitem[{{Hildebrand}(1983)}]{hil83}
{Hildebrand}, R.~H. 1983, \qjras, 24, 267

\bibitem[{{Imanishi} {et~al.}(2007){Imanishi}, {Nakanishi}, {Tamura}, {Oi}, \&
  {Kohno}}]{ima07}
{Imanishi}, M., {Nakanishi}, K., {Tamura}, Y., {Oi}, N., \& {Kohno}, K. 2007,
  \aj, 134, 2366

\bibitem[{{Iono} {et~al.}(2007){Iono}, {Wilson}, {Takakuwa}, {Yun}, {Petitpas},
  {Peck}, {Ho}, {Matsushita}, {Pihlstrom}, \& {Wang}}]{ion07}
{Iono}, D., {Wilson}, C.~D., {Takakuwa}, S., {et~al.} 2007, \apj, 659, 283

\bibitem[{{Klaas} {et~al.}(2001){Klaas}, {Haas}, {M{\"u}ller}, {Chini},
  {Schulz}, {Coulson}, {Hippelein}, {Wilke}, {Albrecht}, \& {Lemke}}]{kla01}
{Klaas}, U., {Haas}, M., {M{\"u}ller}, S.~A.~H., {et~al.} 2001, \aap, 379, 823

\bibitem[{{Komossa} {et~al.}(2003){Komossa}, {Burwitz}, {Hasinger}, {Predehl},
  {Kaastra}, \& {Ikebe}}]{kom03}
{Komossa}, S., {Burwitz}, V., {Hasinger}, G., {et~al.} 2003, \apjl, 582, L15

\bibitem[{{Lahuis} \& {van Dishoeck}(2000)}]{lah00}
{Lahuis}, F., \& {van Dishoeck}, E.~F. 2000, \aap, 355, 699

\bibitem[{{Le Floc'h} {et~al.}(2005){Le Floc'h}, {Papovich}, {Dole}, {Bell},
  {Lagache}, {Rieke}, {Egami}, {P{\'e}rez-Gonz{\'a}lez}, {Alonso-Herrero},
  {Rieke}, {Blaylock}, {Engelbracht}, {Gordon}, {Hines}, {Misselt}, {Morrison},
  \& {Mould}}]{lefl05}
{Le Floc'h}, E., {Papovich}, C., {Dole}, H., {et~al.} 2005, \apj, 632, 169

\bibitem[{{Luhman} {et~al.}(2003){Luhman}, {Satyapal}, {Fischer}, {Wolfire},
  {Sturm}, {Dudley}, {Lutz}, \& {Genzel}}]{luh03}
{Luhman}, M.~L., {Satyapal}, S., {Fischer}, J., {et~al.} 2003, \apj, 594, 758

\bibitem[{{Malhotra} {et~al.}(2001){Malhotra}, {Kaufman}, {Hollenbach},
  {Helou}, {Rubin}, {Brauher}, {Dale}, {Lu}, {Lord}, {Stacey}, {Contursi},
  {Hunter}, \& {Dinerstein}}]{mal01}
{Malhotra}, S., {Kaufman}, M.~J., {Hollenbach}, D., {et~al.} 2001, \apj, 561,
  766

\bibitem[{{Matsushita} {et~al.}(2009){Matsushita}, {Iono}, {Petitpas}, {Chou},
  {Gurwell}, {Hunter}, {Muller}, {Peck}, {Sakamoto}, {Sawada Satoh}, {Wiedner},
  {Wilner}, \& {Wilson}}]{mat09}
{Matsushita}, S., {Iono}, D., {Petitpas}, G.~R., {et~al.} 2009, \apj, 693, 56

\bibitem[{{Max} {et~al.}(2007){Max}, {Canalizo}, \& {de Vries}}]{max07}
{Max}, C.~E., {Canalizo}, G., \& {de Vries}, W.~H. 2007, Science, 316, 1877

\bibitem[{{Medling} {et~al.}(2011){Medling}, {Ammons}, {Max}, {Davies},
  {Engel}, \& {Canalizo}}]{med11}
{Medling}, A.~M., {Ammons}, S.~M., {Max}, C.~E., {et~al.} 2011, \apj, 743, 32

\bibitem[{{Murray} {et~al.}(2005){Murray}, {Quataert}, \& {Thompson}}]{mur05}
{Murray}, N., {Quataert}, E., \& {Thompson}, T.~A. 2005, \apj, 618, 569

\bibitem[{{Nakanishi} {et~al.}(2005){Nakanishi}, {Okumura}, {Kohno}, {Kawabe},
  \& {Nakagawa}}]{nak05}
{Nakanishi}, K., {Okumura}, S.~K., {Kohno}, K., {Kawabe}, R., \& {Nakagawa}, T.
  2005, \pasj, 57, 575

\bibitem[{{Paggi} {et~al.}(2013){Paggi}, {Fabbiano}, {Risaliti}, {Wang}, \&
  {Elvis}}]{pag14}
{Paggi}, A., {Fabbiano}, G., {Risaliti}, G., {Wang}, J., \& {Elvis}, M. 2013,
  ArXiv e-prints, arXiv:1303.2630

\bibitem[{{Paglione} {et~al.}(1995){Paglione}, {Jackson}, {Ishizuki}, \&
  {Rieu}}]{pag95}
{Paglione}, T.~A.~D., {Jackson}, J.~M., {Ishizuki}, S., \& {Rieu}, N.-Q. 1995,
  \aj, 109, 1716

\bibitem[{{Planck Collaboration}(2011{\natexlab{a}})}]{pla11b}
{Planck Collaboration}. 2011{\natexlab{a}}, \aap, 536, A21

\bibitem[{{Planck Collaboration}(2011{\natexlab{b}})}]{pla11a}
---. 2011{\natexlab{b}}, \aap, 536, A25

\bibitem[{{Rangwala} {et~al.}(2011){Rangwala}, {Maloney}, {Glenn}, {Wilson},
  {Rykala}, {Isaak}, {Baes}, {Bendo}, {Boselli}, {Bradford}, {Clements},
  {Cooray}, {Fulton}, {Imhof}, {Kamenetzky}, {Madden}, {Mentuch}, {Sacchi},
  {Sauvage}, {Schirm}, {Smith}, {Spinoglio}, \& {Wolfire}}]{ran11}
{Rangwala}, N., {Maloney}, P.~R., {Glenn}, J., {et~al.} 2011, \apj, 743, 94

\bibitem[{{Sakamoto} {et~al.}(1999){Sakamoto}, {Scoville}, {Yun}, {Crosas},
  {Genzel}, \& {Tacconi}}]{sak99}
{Sakamoto}, K., {Scoville}, N.~Z., {Yun}, M.~S., {et~al.} 1999, \apj, 514, 68

\bibitem[{{Sakamoto} {et~al.}(2008){Sakamoto}, {Wang}, {Wiedner}, {Wang},
  {Peck}, {Zhang}, {Petitpas}, {Ho}, \& {Wilner}}]{sak08}
{Sakamoto}, K., {Wang}, J., {Wiedner}, M.~C., {et~al.} 2008, \apj, 684, 957

\bibitem[{{Sakamoto} {et~al.}(2009){Sakamoto}, {Aalto}, {Wilner}, {Black},
  {Conway}, {Costagliola}, {Peck}, {Spaans}, {Wang}, \& {Wiedner}}]{sak09}
{Sakamoto}, K., {Aalto}, S., {Wilner}, D.~J., {et~al.} 2009, \apjl, 700, L104

\bibitem[{{Sanders} \& {Mirabel}(1996)}]{san96}
{Sanders}, D.~B., \& {Mirabel}, I.~F. 1996, \araa, 34, 749

\bibitem[{{Sanders} {et~al.}(1991){Sanders}, {Scoville}, \& {Soifer}}]{san91}
{Sanders}, D.~B., {Scoville}, N.~Z., \& {Soifer}, B.~T. 1991, \apj, 370, 158

\bibitem[{{Sanders} {et~al.}(1988){Sanders}, {Soifer}, {Elias}, {Madore},
  {Matthews}, {Neugebauer}, \& {Scoville}}]{san88}
{Sanders}, D.~B., {Soifer}, B.~T., {Elias}, J.~H., {et~al.} 1988, \apj, 325, 74

\bibitem[{{Sch{\"o}ier} {et~al.}(2005){Sch{\"o}ier}, {van der Tak}, {van
  Dishoeck}, \& {Black}}]{sch05a}
{Sch{\"o}ier}, F.~L., {van der Tak}, F.~F.~S., {van Dishoeck}, E.~F., \&
  {Black}, J.~H. 2005, \aap, 432, 369

\bibitem[{{Scoville}(2003)}]{sco03}
{Scoville}, N. 2003, Journal of Korean Astronomical Society, 36, 167

\bibitem[{{Scoville} \& {Murchikova}(2013)}]{sco13a}
{Scoville}, N., \& {Murchikova}, L. 2013, \apj, 779, 75

\bibitem[{{Scoville} {et~al.}(2014){Scoville}, {Aussel}, {Sheth}, {Scott},
  {Sanders}, {Ivison}, {Pope}, {Capak}, {Vanden Bout}, {Manohar}, {Kartaltepe},
  {Robertson}, \& {Lilly}}]{sco14}
{Scoville}, N., {Aussel}, H., {Sheth}, K., {et~al.} 2014, \apj, 783, 84

\bibitem[{{Scoville} {et~al.}(2001){Scoville}, {Polletta}, {Ewald}, {Stolovy},
  {Thompson}, \& {Rieke}}]{sco01}
{Scoville}, N.~Z., {Polletta}, M., {Ewald}, S., {et~al.} 2001, \aj, 122, 3017

\bibitem[{{Scoville} \& {Solomon}(1974)}]{sco74}
{Scoville}, N.~Z., \& {Solomon}, P.~M. 1974, \apjl, 187, L67

\bibitem[{{Scoville} {et~al.}(1983){Scoville}, {Young}, \& {Lucy}}]{sco83}
{Scoville}, N.~Z., {Young}, J.~S., \& {Lucy}, L.~B. 1983, \apj, 270, 443

\bibitem[{{Scoville} {et~al.}(1997){Scoville}, {Yun}, \& {Bryant}}]{sco97}
{Scoville}, N.~Z., {Yun}, M.~S., \& {Bryant}, P.~M. 1997, \apj, 484, 702

\bibitem[{{Scoville} {et~al.}(1998){Scoville}, {Evans}, {Dinshaw}, {Thompson},
  {Rieke}, {Schneider}, {Low}, {Hines}, {Stobie}, {Becklin}, \& {Epps}}]{sco98}
{Scoville}, N.~Z., {Evans}, A.~S., {Dinshaw}, N., {et~al.} 1998, \apjl, 492,
  L107+

\bibitem[{{Scoville} {et~al.}(2000){Scoville}, {Evans}, {Thompson}, {Rieke},
  {Hines}, {Low}, {Dinshaw}, {Surace}, \& {Armus}}]{sco00}
{Scoville}, N.~Z., {Evans}, A.~S., {Thompson}, R., {et~al.} 2000, \aj, 119, 991

\bibitem[{{Semenov} {et~al.}(2003){Semenov}, {Henning}, {Helling}, {Ilgner}, \&
  {Sedlmayr}}]{sem03}
{Semenov}, D., {Henning}, T., {Helling}, C., {Ilgner}, M., \& {Sedlmayr}, E.
  2003, \aap, 410, 611

\bibitem[{{Shirley} {et~al.}(2003){Shirley}, {Evans}, {Young}, {Knez}, \&
  {Jaffe}}]{shi03}
{Shirley}, Y.~L., {Evans}, II, N.~J., {Young}, K.~E., {Knez}, C., \& {Jaffe},
  D.~T. 2003, \apjs, 149, 375

\bibitem[{{Spitzer}(1942)}]{spi42}
{Spitzer}, Jr., L. 1942, \apj, 95, 329

\bibitem[{{Stacey} {et~al.}(2010){Stacey}, {Hailey-Dunsheath}, {Ferkinhoff},
  {Nikola}, {Parshley}, {Benford}, {Staguhn}, \& {Fiolet}}]{sta10}
{Stacey}, G.~J., {Hailey-Dunsheath}, S., {Ferkinhoff}, C., {et~al.} 2010, \apj,
  724, 957

\bibitem[{{Tacconi} {et~al.}(1999){Tacconi}, {Genzel}, {Tecza}, {Gallimore},
  {Downes}, \& {Scoville}}]{tac99}
{Tacconi}, L.~J., {Genzel}, R., {Tecza}, M., {et~al.} 1999, \apj, 524, 732

\bibitem[{{Tecza} {et~al.}(2000){Tecza}, {Genzel}, {Tacconi}, {Anders},
  {Tacconi-Garman}, \& {Thatte}}]{tec00}
{Tecza}, M., {Genzel}, R., {Tacconi}, L.~J., {et~al.} 2000, \apj, 537, 178

\bibitem[{{Thompson} {et~al.}(2005){Thompson}, {Quataert}, \& {Murray}}]{tho05}
{Thompson}, T.~A., {Quataert}, E., \& {Murray}, N. 2005, \apj, 630, 167

\bibitem[{{Wilson} {et~al.}(2014){Wilson}, {Rangwala}, {Glenn}, {Maloney},
  {Spinoglio}, \& {Pereira-Santaella}}]{wil14}
{Wilson}, C.~D., {Rangwala}, N., {Glenn}, J., {et~al.} 2014, \apjl, 789, L36

\end{thebibliography}


\appendix 

\section{Kinematic Modeling and Deconvolution}\label{app}


We make use of the kinematic deconvolution technique developed by \cite{sco83}. This yields a maximum likelihood solution for the disk model 
with axisymmetric emissivity distribution, rotation curve and velocity dispersion which yields line profiles best matching the observed 
line profiles, given the instrumental spatial and kinematic resolutions. 
The technique is analogous to Doppler radar of rotating planets and in our earlier 
application to the CO single dish observations of NGC 1068, it revealed a molecular ring at $\sim12$\arcsec ~radius although the single dish beam was 60\arcsec ~resolution \citep{sco83}. 


We used the same algorithm described in \cite{sco83}, with the addition of an external computation loop, varying rotation curve parameters to minimize the 
overall $\chi^2$ of the observed line profiles compared with those 'observed' from the model using the observational resolution parameters. The rotation curve was parametrized by the simple 3-part function:
 \begin{eqnarray}
 V(R) &=& V_0 {R \over{R_0}}~ \kms  ~\rm at ~R \leq R_0 \nonumber  \\
  V(R) &=& V_0   ~ \kms  ~\rm at ~ R_0 < R < 3R_0 \nonumber\\
  V(R) &=& V_0 \left({3R_0  \over{R}}\right)^{1/2} ~ \kms ~\rm at ~R \geq 3R_0   
\end{eqnarray}
\noindent where $V_0$ and $R_0$ were parameters varied to minimize the $\chi^2$. The exact form of the rotation curve is somewhat arbitrary -- however, it is reasonable to have a rising portion ($R \leq R_0$) and then a fairly flat  portion to mimic observed galactic rotation curves. The last segment, a Keplerian 
falloff, is physically motivated by the finite radial extent of the massive nuclear disks. 

The remaining parameters are: the velocity dispersion $\sigma_v$ 
of the gas, the inclination i of the disk, the PA of the major axis of the disk and the emissivity as a function of radius. The latter was taken to be a step function in radius 
with 30 equal width bins in R with no imposed continuity between adjacent radial bins. Positions and velocities were all measured relative to the spatial and velocity 
centroid of the HCN emission; i.e. the centroid position and velocity is adopted as the nucleus (R=0) and the systemic velocity of each disk. (These offsets 
were in all cases $\leq 0.05$\arcsec~ and $\leq 20$ \kms from the dust continuum nuclear peaks and the systemic velocities.)

\end{document}